\documentclass[sigconf]{acmart}
\usepackage{xcolor}
\usepackage{colortbl}

\usepackage[]{hyperref}
\usepackage{braket}
\usepackage{microtype}[final]
\usepackage{textcomp}
\usepackage{multirow}
\usepackage{makecell}
\usepackage{framed}

\usepackage{subcaption}
\usepackage{algorithmic}

\definecolor{lightgray}{gray}{0.95}
\definecolor{teal}{rgb}{0.0, 0.5, 0.5}
\definecolor{amber}{rgb}{1.0, 0.49, 0.0}
\definecolor{purple}{rgb}{0.6, 0.4, 0.8}
\definecolor{blue-violet}{rgb}{0.54, 0.17, 0.89}

\newcommand{\ours}{{\sc{gladiator }}}
\newcommand{\ournspace}{{\sc{gladiator}}}
\newcommand{\oursm}{{\sc{gladiator+m }}}
\newcommand{\oursmnspace}{{\sc{gladiator+m}}}
\newcommand{\oursd}{{\sc{gladiator-d }}}
\newcommand{\oursdnspace}{{\sc{gladiator-d}}}
\newcommand{\oursdm}{{\sc{gladiator-d+m }}}
\newcommand{\oursdmnspace}{{\sc{gladiator-d+m}}}
\newcommand{\eraser}{{\sc{eraser }}}
\newcommand{\eraserm}{{\sc{eraser+m }}}
\newcommand{\erasermnspace}{{\sc{eraser+m}}}
\newcommand{\erasernspace}{{\sc{eraser}}}

\newcommand{\rev}[1]{\textcolor{black}{#1}}

\usepackage[linesnumbered,ruled]{algorithm2e}
\AtBeginDocument{%
  }

\usepackage{framed}
\definecolor{shadecolor}{rgb}{0.95,0.95,0.95}

\newenvironment{highlightitembox}{%
    \begin{shaded}%
}{%
    \end{shaded}%
}

\copyrightyear{2025}
\acmYear{2025}
\setcopyright{cc}
\setcctype{by-nd}
\acmConference[MICRO '25]{58th IEEE/ACM International Symposium on Microarchitecture}{October 18--22, 2025}{Seoul, Republic of Korea}
\acmBooktitle{58th IEEE/ACM International Symposium on Microarchitecture (MICRO '25), October 18--22, 2025, Seoul, Republic of Korea}\acmDOI{10.1145/3725843.3756053}
\acmISBN{979-8-4007-1573-0/2025/10}

\begin{document}

%

\title{Accurate Leakage Speculation for Quantum Error Correction}

\author{Chaithanya Naik Mude}
\orcid{0009-0002-0531-3551}
\affiliation{%
  \institution{University of Wisconsin-Madison}
  \city{Madison}
  \state{WI}
  \country{USA}
}
\email{cmude@wisc.edu}

\author{Swamit Tannu}
\orcid{0000-0003-4479-7413}
\affiliation{%
  \institution{University of Wisconsin-Madison}
  \city{Madison}
  \state{WI}
  \country{USA}
}
\email{swamit@cs.wisc.edu}


    


\begin{abstract}
{\color{black}

Quantum Error Correction (QEC) protects qubits against bit- and phase-flip errors in the $\ket{0}/\ket{1}$ subspace, but physical qubits can also leak into higher energy levels (e.g., $\ket{2}$). Leakage is especially harmful, as it corrupts all subsequent syndrome measurements and can spread to neighboring qubits. Detecting leakage on data qubits is particularly challenging, since they are never measured directly during QEC cycles. Prior work, such as \textsc{eraser}~\cite{ERASER}, addresses this by inferring leakage from syndrome patterns using a fixed heuristic. However, this approach often misclassifies benign syndromes, triggering excessive leakage-reduction circuits (LRCs). Because LRCs are themselves noisy and slow, these false triggers lengthen QEC cycles and inflate logical error rates.

We propose \textsc{gladiator}, a general and adaptable leakage speculation framework that works across surface code, color code, and qLDPC codes. Offline, \textsc{gladiator} builds a code-aware error-propagation graph calibrated to device data. Online, it classifies each syndrome in a few nanoseconds and schedules LRC only when the observed pattern is provably leakage-dominated. This precise speculation eliminates up to $3\times$ (and on average $2\times$) unnecessary LRCs, shortens QEC cycles, and suppresses false positives at their source. Evaluated on standard fault-tolerant benchmarks, \textsc{gladiator} delivers $1.7\times$–$3.9\times$ speedups and 16\% reduction in logical error rate, advancing the efficiency of fault-tolerant quantum computing.

}
\end{abstract}






\maketitle
\section{Introduction}

Quantum computers promise significant speedups over classical computers for quantum simulations, unstructured search, and factorization but are limited by their sensitivity to errors, as quantum information is delicate. The qubits used to store quantum information are highly error-prone, limiting the potential applications of quantum computing. Quantum error correction (QEC) codes can be leveraged to detect and correct the physical errors to enable fault-tolerant memory and computation. In QEC, a large number of noisy physical qubits are used to encode the quantum information into logical qubits, which can detect and correct errors.

Albeit at small scales, many industry and academic labs~\cite{logical_processor_NA, xu2023constantoverhead_NA, LDPC_ibm, acharya2024quantumerrorcorrectionsurface} are actively demonstrating the efficacy of different QEC protocols on a variety of quantum hardware platforms. Some of these experimental demonstrations, especially on superconducting architectures~\cite{experimental_QEC_SC, DQLR}, are revealing significant challenges in enabling effective quantum error correction. Leakage errors, for instance, pose a significant challenge to enabling effective QEC protocols. 
 
QEC protocols, while effective for errors within the standard computational basis of qubits ($\ket{0}, \ket{1}$), can not detect errors due to leaked states ($\ket{L}$), where qubits occupy higher energy levels ($\ket{2}, \ket{3}$, etc.). Such leakage extends beyond intended computational states, posing significant challenges to QEC's effectiveness in error correction as the leaked qubits result in malfunctioning gates corrupting syndrome generation during QEC, leading to faulty detection and correction. Recent experimental studies by IBM~\cite{leakage_supression_IBM} and Google QuantumAI~\cite{experimental_QEC_SC, acharya2024quantumerrorcorrectionsurface, sycamore_leakage} highlight this issue, showing leakage errors limiting QEC codes' ability to detect and correct errors.

To improve the effectiveness of QEC at scale, we need techniques to detect and remove the correlated error caused by leakage, especially on data qubits, which are not measured directly and known to accumulate leakage errors~\cite{sycamore_leakage, DQLR, leakage_elimination_weakly_non_linear}. By executing additional operations known as Leakage Reduction Circuit (LRC) gadgets, leaked qubits are forced to return to the computational subspace. Prior works \cite{NC_critical_faults, Battistel_2021, lacroix2023fast, Marques_2023, leakage_elimination_weakly_non_linear, Hayes_2020, FTQC_local_leakage} have focused on device-specific mitigation strategies requiring specialized hardware for implementing LRCs, while other methods use more general-purpose gates such as SWAP gates to implement LRCs~\cite{McEwen_2021, Brown_2019}.

\begin{figure*}[ht]
    \centering
    \includegraphics[width=0.9\linewidth]{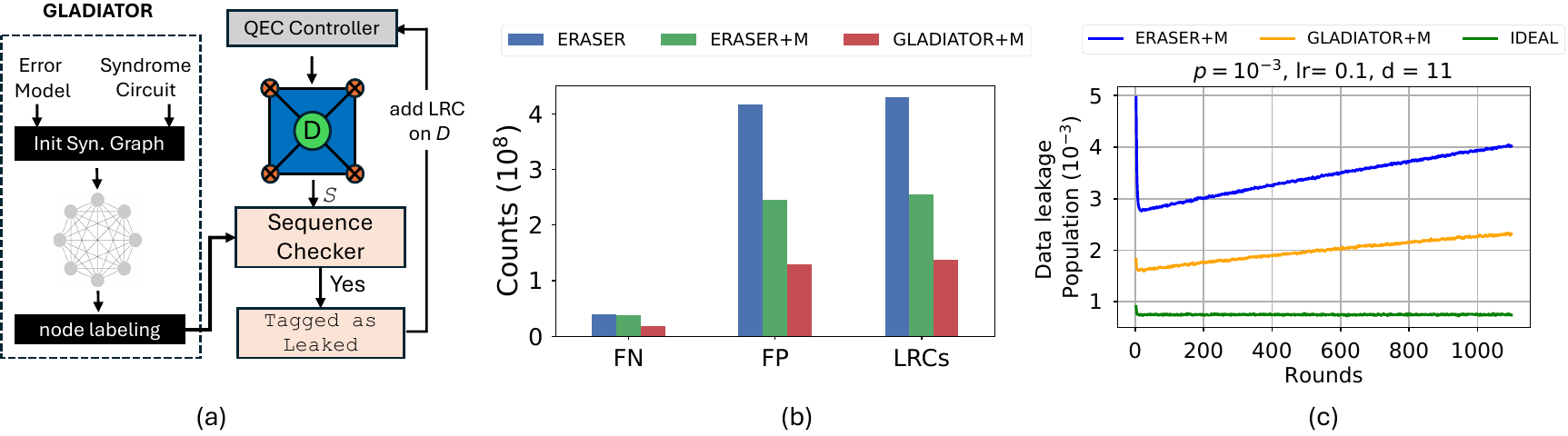}
    \caption{(a) Proposed \ours Design. (b) Comparing the effectiveness of \ours and \eraser using False Negative (FN), False Positive (FP), and LRC utilization. (c) The leakage population for surface code with code distance d =11, when executed for $100d$ QEC rounds, assuming the probability of non-leakage error $p_{e}=10^{-3}$ and probability of leakage is $p_{leak} = lr\times p_{e} = 10^{-4}$. }
    \label{fig:introduction}
    \Description{Proposed design of GLADIATOR along with the improvements in LRC usage and reduction in leakage population.}
\end{figure*}

Though these LRC operations significantly reduce leakage effects, they increase QEC's cycle latency and resource overhead. For example, a recent work, \eraser~\cite{ERASER}, highlighted the inefficiencies in unconditional applications of LRCs and developed a speculation-based leakage detection technique using syndrome measurements to insert LRCs judiciously, as adding LRCs constitutes doing additional physical operations, increasing the gate errors and even increasing the probability of leakage itself. \eraser observes that when a data qubit is leaked, it manifests as a syndrome pattern where atleast 50\% of the parity qubits are flipped. Whenever such a pattern is encountered, \eraser speculatively applies the LRCs. Furthermore, \eraser uses a {\em Multi Level Readout (MLR)} on all parity qubits to detect leaked parity qubits and trigger the application of LRCs. However, we observe several limitations of \erasernspace. 


{\textbf{High False Positive Rate.}} \erasernspace's strategy leads to a significant rate of false positives due to its assumption that 50\% of bit flips would mainly arise from leakage. In reality, many of these bit flips result from other types of errors occurring in computational basis. This high false positive rate degrades reliability, adding unnecessary LRCs introducing additional gate and leakage errors.

{\textbf{Growing Leakage Population.}} Accumulated leakage can severely impair QEC, potentially making it ineffective because leakage paralyzes syndrome measurement \cite{coping_qubit_leakage}. Ideally, a leakage speculation policy should stabilize or reduce leakage levels over time. Due to imperfection in qubit devices, the insertion of unnecessary LRCs can increase leakage. Although \eraser effectively reduces leakage by strategically applying LRCs, our tests over 100 QEC cycles reveal a continuous rise in leakage population when \eraser is used. This increase is primarily due to \erasernspace's high rate of false positives, which mistakenly flag qubits as leaky, leading to unnecessary LRC insertions. The speculation strategy \eraser uses is rigid and ineffective for long-running QEC cycles, which need to withstand variations in qubit characteristics, and leakage accumulation.

{\color{black}

\textbf{Generalizability} is essential for scalable leakage mitigation. \erasernspace's heuristic-based approach exploits the symmetry and regularity of the surface code, where each data qubit interacts with four ancilla qubits, yielding rich syndrome information. However, {\color{black} ERASER's strategy does not generalize to other QEC codes, such as color codes or quantum LDPC codes, where syndrome information per data qubit is sparse, and the stabilizer connectivity is irregular and often asymmetric}. In these settings, accurately inferring leakage becomes significantly harder due to the limited and noisy observables. The challenge is exacerbated in high-rate codes, where each syndrome bit is influenced by fewer qubits. Furthermore, alternative approaches like walking surface code~\cite{eickbusch2024demonstratingdynamicsurfacecodes} rely on planar, cyclic role exchanges that are incompatible with non-planar topologies and irregular interaction graphs required by codes like qLDPC or hypergraph product codes (HGP) \cite{kang2025quitsmodularqldpccode}.}

{ \textbf{GLADIATOR.}} We propose a general design framework, \ournspace, that can guide the leakage speculation strategy by using the error model and error correcting code to estimate syndromes that are most likely caused by the data qubit leakage.  
We observe that the patterns \eraser flags as leakage may also result from non-leakage errors caused by imperfect gates and measurements, which can approximately be ten times more likely than leakage. Our characterization of IBM hardware shows that with leakage on the data qubit, the CNOT gates malfunction, producing independent random bit flips; therefore, certain syndrome patterns are more likely to be caused by leakage-driven random bit flips than consistent bit flips caused by non-leakage errors. For example, pattern ``0011'' is more likely to be caused by non-leakage (i.e, data qubit excitation), while the pattern ``1001'' most likely indicates a leakage. 

As shown in Figure~\ref{fig:introduction}(a), we formulate the leakage diagnosis using syndrome patterns as a graph labeling problem. First, we construct a leakage graph in which a node represents a 4-bit syndrome pattern associated with a data qubit measured during the surface code syndrome generation cycle. The nodes are connected via directed and weighted edges. The directionality of edges indicates the possibility of a syndrome pattern in the source node transforming into a target pattern due to leakage, with the transformation likelihood determining the weight of these edges. Similarly, we build the syndrome transformation graph due to non-leakage errors, where we only account for transformation results from non-leakage errors. We merge both the non-leakage and leakage graphs so that the edges are adjusted to reflect the combined probabilities of these transitions. Using relative transition rates, we label syndrome nodes as leakage or non-leakage nodes. Our graph-based approach allows for the systematic construction of a diagnosis framework, where nodes are labeled based on their probability of representing leakage or non-leakage scenarios, using experimental data to guide the decision-making process. Under identical error configurations, \eraser flags \textit{11/16} syndrome patterns as leakage-causing patterns, whereas \ours flags only \textit{8/16} as the most likely leakage patterns. This reduces the false positive rate by $1.91\times$ as shown in Figure~\ref{fig:introduction}(b), thereby reducing the data leakage population by $1.73\times$ as shown in Figure~\ref{fig:introduction}(c), for code distance $d=11$ in $100d$ rounds.

{\textbf{Speculation Inaccuracy.}} We decrease the total counts of false positives and false negatives combined by around $3.11\times$ for \ours over \erasernspace. For the MLR-enhanced version of \erasernspace, i.e, \eraserm around $1.92\times$ with \oursm for d=11.

{\textbf{Leakage Population.}} \oursm produces $1.73\times$ less data leakages compared to \erasermnspace, after $100$ QEC cycles for $d=11$.

{\textbf{Resource Overheads}} Our method uses less than 0.1\% LUTs at even code distance 25. Compared to \textbf{\erasernspace}, it achieves at least $17\times$ reduction in FPGA resource usage across code distances 5–25. Our generalizable approach ensures principled and robust leakage mitigation strategies. The key differences between the prior work \eraserm and our strategies, \textsc{staggering} and \oursmnspace, primarily lie in their approach to leverage syndrome-based speculation, and Multi-Level Readout (MLR) as summarized in Table~\ref{tab:differences_in_policies}.


\rev{
\begin{table}[h]
\begin{center}
\small 
\centering
\caption{\rev{Leakage Mitigation: \textbf{\ournspace} vs. Prior Work 
}}\label{Policy_diff}
\vspace{-0.1in}
\label{tab:differences_in_policies}
\setlength{\tabcolsep}{0.06cm}
\renewcommand{\arraystretch}{1.15}
\rev{
\begin{tabular}{c c c c c c} 
\toprule
Method & Speculation & Rounds & MLR & Adaptability\\
\midrule 
Always-LRC & No & 0 & No & No \\
\textsc{staggering (ours)} & No & 0 & No & No \\
DQLR\cite{DQLR} & No & 0 & Yes & No \\
\textbf{\eraserm}\cite{ERASER} & Yes & 1 & Yes & No \\
\textbf{\oursm} \textsc{(ours)} & Yes & 1 & Yes & Yes \\
\textbf{\oursdm} \textsc{(ours)} & Yes & 2 & Yes & Yes \\
\bottomrule
\end{tabular} }
\end{center}
\end{table}
 } 
\section{Background and Motivation}

\subsection{Noisy Qubits and Leakage Errors}

\begin{figure}[b]
    \centering
    \includegraphics[width=\linewidth]{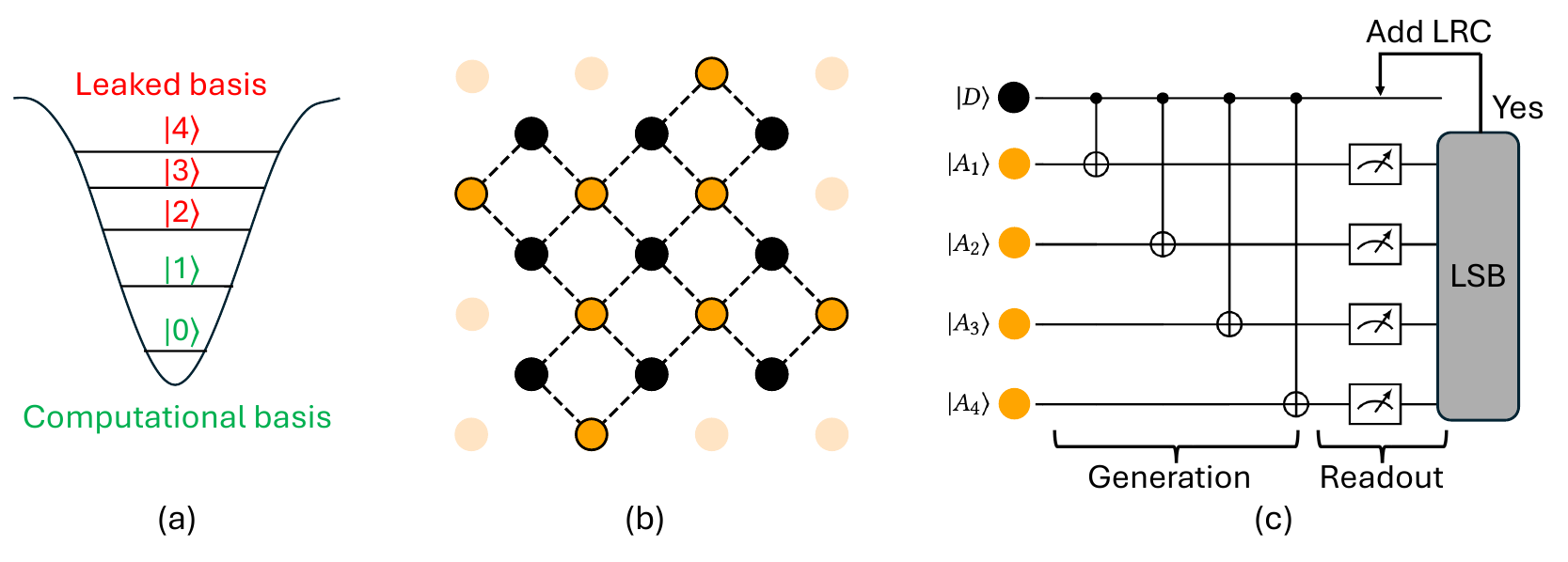}
    \caption {(a) Qubit device with computational basis corresponding to $\ket{0}$,$\ket{1}$ and leaked basis corresponds to higher energy quantum levels. (b) Surface code of distance 3, with black qubits corresponds to data qubits and others corresponding to the parity qubits. (c) Syndrome Generation for LRC insertion using Leakage Speculation Block (LSB).}
    \label{fig:back}
    \Description{Leakage levels in the qubit and overview of surface code of code-distance 3 and speculative insertion of LRC}
\end{figure}

Qubits, by design, are anharmonic oscillators intended to function as two-level systems suitable for quantum computing.  Unfortunately, they inherently possess multiple energy levels, complicating their operation within the desired computational subspace. As shown in Figure~\ref{fig:back}(a), this subspace typically comprises the two lowest energy states, labeled as $\ket{0}$ and $\ket{1}$. Qubits are highly sensitive to errors and can be broadly categorized into two types:

\textbf{Intra-subspace errors.} These occur within the computational subspace itself. Errors of this type involve transitions between the $\ket{0}$ and $\ket{1}$ states or create superpositions of these states and can be corrected by standard quantum error correction (QEC) techniques.

\textbf{Leakage errors.} Unlike intra-subspace errors, leakage errors involve transitions of a qubit's state to energy levels outside of the designated computational subspace, such as $\ket{2}$, $\ket{3}$, or higher. These errors effectively ``hide" the qubit state from the standard error correction mechanisms. Addressing leakage requires specialized techniques to detect when a qubit exits the computational subspace and implement strategies to return it to the correct operational state. This might involve additional leakage reduction circuits, specifically designed to monitor and correct leakage. Other strategies include the integration of qudit-error correcting codes~\cite{PhysRevA.103.042420,Looi_2008,PhysRevLett.131.200602}.

\subsection{Impact of Leakage on QEC} 

Quantum error-correcting codes reduce error rates by encoding a logical qubit across many physical qubits and continuously measuring parity checks. Surface codes is a leading approach, using $2d^2 - 1$ qubits for a distance‑$d$ code with $d^2$ data qubits and $d^2 - 1$ parity qubits. Data qubits are entangled with parity qubits via stabilizer circuits to detect and correct bit-flip (X), phase-flip (Z), or combined errors. A distance‑$d$ code can correct up to $\frac{d - 1}{2}$ errors.

Leakage errors excite qubits to higher energy states. In this scenario, the traditional syndrome extraction circuits used in QEC fail to detect the leakage, as their design does not account for these leaked states. Therefore, even though the syndrome extraction circuits are run multiple times, the iterative process cannot identify or correct leakage errors, resulting in the accumulation of leakage errors and catastrophic failures~\cite{leakage_toric_code, NC_critical_faults, coping_qubit_leakage}.

Furthermore, the two-qubit gates such as CZ and CX, used in the syndrome generator circuits malfunction in the presence of leakage \cite{SC_leak_effect_QEC}, further complicating the error correction process. This malfunctioning alters the expected behavior of the gates, leading to incorrect or incomplete syndrome generation. Thus, without modifications or additional mechanisms specifically targeting leakage, standard QEC methods remain ineffective against leakage errors, as detailed in Section~\ref{LeakageIBM}, where we will discuss leakage injection experiments conducted on IBM systems. To mitigate the impact of leakage on QEC, the \eraserm framework employs a leakage speculation block (LSB) to analyze syndrome measurements, detect leakage, and apply leakage reduction circuits to restore the qubit to one of its computational states, as illustrated in Figure ~\ref{fig:back}(c).

\subsection{Leakage Characterization on IBM Systems} \label{LeakageIBM}

QEC depends on reliable gate operations. However, leaked qubits disrupt gate behavior, undermining QEC efficacy. Prior work~\cite{sycamore_leakage, Cai2019siliconsurfacecode} has shown that QEC performance is tightly coupled to the hardware’s error profile. In superconducting qubits, leakage is rare ($\sim$$10^{-3}$) and stochastic~\cite{NC_critical_faults}, making systematic analysis difficult.

\begin{figure}[ht]
    \centering
\includegraphics[width=\linewidth]{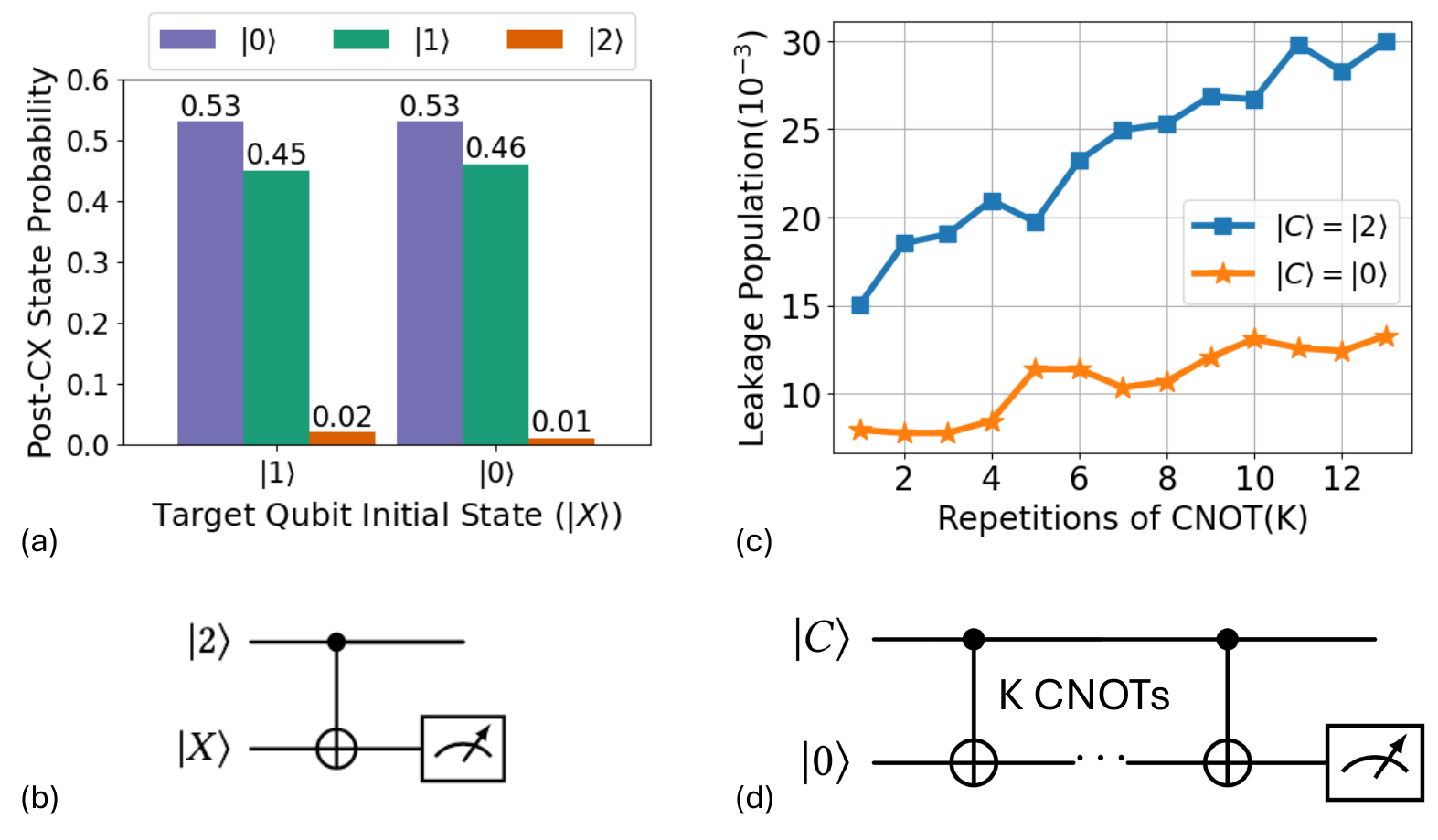}
    \caption{ {\color{black} Leakage Injection Experiment on IBM hardware- (a) The probability of the measured state when circuit (b), a single physical CNOT with one of the qubits leaked, is executed. (c) The leakage population of measured qubit when circuit. (d) with one physical CNOT is repeatedly executed. We ran 10,000 shots using Qiskit Pulse~\cite{Qiskit, IBMGuide}.}}
\label{fig:CX_leakage_IBM}
\Description{Leakage accumulation and CNOT gate malfunction experiments on IBM hardware}
\end{figure}

To study leakage-induced faults in syndrome generation, we conduct experiments on IBM quantum systems using a targeted leakage injection method. Specifically, we initialize qubits in the $\ket{2}$ (leaked) state and repeatedly run syndrome generation circuits. This setup lets us observe how leakage propagates between data and parity qubits and track leakage dynamics over multiple QEC cycles. Due to hardware constraints, we focus on the core component of syndrome circuits—the CNOT gate. Figure~\ref{fig:CX_leakage_IBM}(a) shows when the control qubit is leaked, the target qubit toggles between $\ket{0}$ and $\ket{1}$, effectively producing a mixed state, inducing a 50\% bit-flip error. 
\begin{center}
    \textit{Leakage disrupts CNOT operations and must be mitigated for QEC to remain effective.} 
\end{center}

To examine how leakage accumulates, we repeat CNOT operations with and without injecting leakage. This emulates multiple QEC rounds. As shown in Fig.~\ref{fig:CX_leakage_IBM}(c), leakage population grows over time when initialized, and remains low without injection, highlighting how a single leaked qubit can spread leakage.

{\color{black}
\subsection{Taxonomy of Leakage Reduction Circuits}

Leakage Reduction Circuits (LRCs) are physical operations that convert leakage errors into Pauli errors that QEC can detect. LRCs fall into four categories: \textit{reset-based}, \textit{specialized hardware}, \textit{qubit role exchange}, and \textit{leakage-to-erasure conversion}.

\textbf{Reset-Based LRCs.} These approaches use SWAP gates or Reset protocols~\cite{McEwen_2021} to offload leakage to ancilla qubits for conditional reset. While hardware-friendly, they introduce circuit depth overhead and depend on high-fidelity leakage detection; misidentification can exacerbate logical error rates (LER).

\textbf{Specialized Hardware LRCs.} Techniques like Data Qubit Leakage Removal (DQLR)~\cite{DQLR} coherently transfer leakage population via a Leakage-iSWAP gate to a fast-reset qubit. Although effective, DQLR requires precise control, custom pulses, and complex calibration, and remains non-fault-tolerant.

\textbf{Qubit Role Exchange.} The walking surface code~\cite{eickbusch2024demonstratingdynamicsurfacecodes} alternates data and ancilla roles across QEC rounds, enabling staggered resets. While this avoids specialized hardware, it incurs time overhead and {\color{black}\textit{does not generalize well beyond surface codes.}}

\textbf{Leakage-to-Erasure Conversion.} Some platforms convert leakage into detectable erasures by encoding in metastable subspaces or using fluorescence-based detection~\cite{Wu_2022, Dual-RailQubit}. Despite high conversion rates, residual leakage and imperfect resets persist.

While LRC techniques vary in mechanism and overhead, they are often applied naively in every QEC round. However, LRCs are \emph{not fault-tolerant}, as excessive application introduces gate and leakage errors, while infrequent application allows leakage to accumulate. To be effective, LRC scheduling must satisfy three key criteria: 

\begin{highlightitembox}
\begin{itemize}
    \item \textcolor{black}{\textbf{Accuracy:} Minimize false positives (unnecessary resets) and false negatives (missed leakage).}
    \item \textcolor{black}{\textbf{Generalizability:} Maintain efficacy across QEC codes (e.g., surface codes, color codes, HGP, BPC codes).}
    \item \textcolor{black}{\textbf{Adaptability:} Remain effective under variable noise models and hardware yield variations~\cite{scaling_chiplet_Smith, codesign_QEC_chiplets}.}
\end{itemize}
\end{highlightitembox}
}

{\color{black}
\section{Limitations of Prior Work }

\subsection{Open-Loop vs. Closed-Loop LRC Scheduling}

Leakage mitigation strategies fundamentally differ in how they decide when to apply Leakage Reduction Circuits (LRCs). We categorize these approaches into two paradigms:

\textbf{Open-Loop Scheduling} applies LRCs at predetermined intervals without detecting actual leakage events. These methods are hardware-friendly and require minimal real-time decision-making, but risk applying unnecessary LRCs when no leakage is present.

\textbf{Closed-Loop Scheduling} applies LRCs based on inferred leakage from syndrome measurements. While offering higher precision by targeting actual leakage events, these methods require reliable leakage detection and low-latency control decisions.

The key trade-off lies between simplicity and precision, as open-loop methods execute a large number of unnecessary LRCs, while closed-loop methods risk missing leakage events or making incorrect inferences from noisy syndrome data.

\subsection{ERASER: Motivating Closed-Loop Control}

\textsc{eraser}~\cite{ERASER} introduced the first closed-loop leakage mitigation framework, demonstrating significant improvements over naive open-loop baselines. The core insight was that leaked qubits corrupt syndrome measurements in predictable patterns, enabling leakage inference from stabilizer measurement anomalies.

\textbf{ERASER's Detection Strategy}: \textsc{eraser} infers data qubit leakage by monitoring downstream effects on syndrome measurements. A leaked data qubit corrupts CNOT gate behavior, producing random measurement outcomes that typically cause approximately 50\% of connected syndrome bits to flip. \eraser triggers LRCs when either: (1) 50\% or more of connected syndrome bits flip, or (2) Multi-Level Readout (MLR) on parity qubits directly detects leakage.

\textbf{Baseline Comparison}: \textsc{eraser} demonstrated its effectiveness against \textit{Always-LRC}, an open-loop policy that applies LRCs to all qubits simultaneously after every QEC round. This comparison showed substantial improvements in logical error rates, establishing the value of selective LRC application.

\subsection{ERASER's Design Limitations}

Our analysis reveals two critical limitations in \textsc{eraser}'s approach that motivate the need for more sophisticated closed-loop methods:

\textbf{Limitation 1: High False Positive Rate}: \textsc{eraser}'s rigid 50\% threshold assumption leads to frequent misclassification of standard gate error patterns as leakage events. Table~\ref{tab:efficacy_fp_fn} shows that many syndrome patterns attributed to leakage are more likely caused by non-leakage errors, resulting in unnecessary LRC applications that can introduce additional operational and leakage errors.

\textbf{Limitation 2: Inadequate Leakage Population Control}: The high false positive rate prevents effective leakage population management. Excessive LRC applications not only waste resources but can paradoxically increase leakage population over time due to LRC-induced errors, as evidenced by the gradual increase in leakage levels shown in Table~\ref{tab:efficacy_fp_fn}.

{\color{black}\textbf{Limitation 3: Code-Specific Heuristics and Poor Generalization}}: \textsc{eraser}'s detection strategy relies on code-specific assumptions that limit its applicability beyond surface codes. The 50\% threshold heuristic exploits the symmetric structure of surface codes, where data qubits are typically connected to 3-4 ancilla qubits, making the ``half-flip'' pattern a reasonable indicator of leakage-induced CNOT corruption.

However, this heuristic fails to generalize to other quantum error correction codes with different connectivity patterns and structural properties. In color codes, for example, data qubits are connected to only 2-3 ancilla qubits, significantly altering the expected syndrome patterns from leaked qubits. Similarly, qLDPC codes exhibit irregular connectivity graphs that break the symmetry assumptions underlying \textsc{eraser}'s detection logic.

Our experimental analysis reveals that when applied to color codes, \textsc{eraser}'s heuristic becomes overly sensitive, triggering LRCs for almost all error patterns, therefore, effectively reducing to the \textit{Always-LRC} baseline that \textsc{eraser} originally demonstrated to be suboptimal. This poor generalization severely limits \textsc{eraser}'s practical applicability as the quantum computing field explores diverse QEC code families beyond surface codes.

These three fundamental limitations demonstrate the need for more sophisticated approaches that can achieve \textsc{eraser}'s selective application benefits while dramatically reducing false positive rates and generalizing across diverse QEC code structures.

\renewcommand{\arraystretch}{1.15}
\begin{table}[t] 
\centering
\caption{Leakage Detection Efficacy of {\sc eraser}({\sc er})} 
\label{tab:efficacy_fp_fn}
\setlength{\tabcolsep}{1pt} 
\small
\begin{tabular}{@{}lcccccc@{}} 
\toprule
Metrics & \makecell{Always LRC} & \makecell{ER} & \makecell{ER+M} & \makecell{M} & \makecell{Staggered} & \makecell{Ours}\\ \midrule 
FN         & $1$ & $3.9$ & $3.8$ & $6.3$ & $1.6$ & $3.2$\\
FP         & $1$ & $0.06$ & $0.04$ & $0.05$ & $0.5$ &$0.021$\\
LRCs & $1$              & $0.06$                & $0.04$               & $0.05$ & $0.5$ & $0.022$\\
\midrule
\rowcolor{blue!15}\textbf{Leak-70$(10^{-3})$} & $4.36$              & $4.19$                & $2.97$               & $6.01$ & $4.87$ & $3.24$\\
\midrule
\rowcolor{red!15}\textbf{Leak-700$(10^{-3})$} & $6.73$              & $4.94$                & $3.78$               & $6.87$ & $4.54$& $3.35$\\
\bottomrule
\end{tabular}
\end{table}

\subsection{Gaps in ERASER's Analysis}

While \textsc{eraser}'s closed-loop approach represents a significant advance, its experimental evaluation has notable gaps that limit our understanding of the open-loop vs. closed-loop trade-offs:

\textbf{Weak Open-Loop Baseline}: \textsc{eraser} only compared against \textit{Always-LRC}, which simultaneously resets all qubits---a policy known to increase correlated errors and gate overhead. This naive baseline does not represent the best possible open-loop strategy.

\textbf{Missing Structured Open-Loop Analysis}: \textsc{eraser} did not explore whether more intelligent open-loop scheduling could narrow the performance gap. Specifically, a spatially staggered LRC application could reduce correlated errors while maintaining simplicity.

\textbf{Incomplete MLR-only Evaluation}: The original \textsc{eraser} work did not evaluate against MLR-only detection (using only parity qubit measurements), making it difficult to assess the value of syndrome-based inference. To address these analytical gaps, we study the Staggered application of LRCs with MLR.

\begin{figure}[ht]
    \centering
\includegraphics[width=\linewidth]{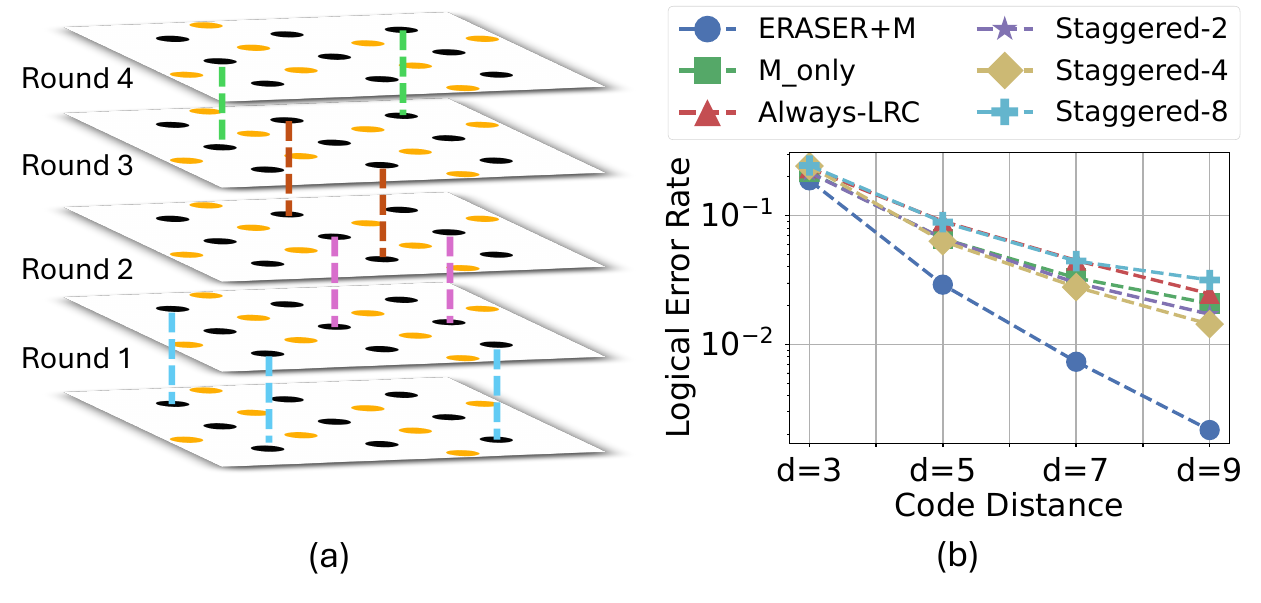}
    \caption{(a) Qubit coloring (Dotted lines as color groups) for round-robin LRC scheduling in Staggered-LRC. (b) Logical Error Rate (LER) across open-loop policies and ERASER+M.}

\label{fig:stagger_LRC}
\Description {Staggered Design}
\end{figure}

\subsection{Reducing Leakage via Staggered LRCs}\label{sec:speculation}

We propose \textit{Staggered Always-LRC}, a structured open-loop policy that schedules LRCs as an $n$-coloring problem on the qubit interaction graph. No adjacent or diagonally neighboring qubits share the same color, and each color group is reset in a round-robin fashion across QEC rounds as shown in Figure~\ref{fig:stagger_LRC}(a) with color groups represented by dotted lines. This spatial staggering minimizes correlated faults while maintaining the simplicity of open-loop control. While Staggered Always-LRC does not universally outperform \erasernspace, it significantly narrows the gap in low leakage mobility scenarios. At low mobility $(\approx1\%)$, the LER gap between Staggered Always-LRC and \eraser shrinks by $2\times$ compared to high mobility $(\approx10\%)$.

\textbf{Key Insight}: When leakage mobility, which is the probability of leakage transport between qubits is low, and LRC gate errors are modest, structured open-loop strategies can be competitive with closed-loop methods, especially at smaller code distances.

While improved open-loop scheduling narrows the gap, closed-loop methods still offer advantages in scenarios with higher leakage rates or mobility. However, \textsc{eraser}'s heuristic-based approach suffers from high false positive rates.

We introduce \textsc{gladiator}, a principled closed-loop framework that models leakage speculation as a graph labeling problem. Rather than relying on simple bit-flip-count-based heuristics, \textsc{gladiator} uses probabilistic inference to distinguish leakage-induced syndrome patterns from those caused by standard gate errors.

\section{GLADIATOR}

To address the limitations of \erasernspace, we propose a more accurate, adaptable, and generalizable approach to leakage speculation: {\em Graphical Model for Leakage Detection in Syndrome Generator ({\sc gladiator})}. This section outlines {\sc gladiator}'s key insights and design.

\subsection{Not All Bit-flips are caused by Leakage}

A leakage error in a data qubit can induce specific bit-flip patterns in the measured syndrome.  \eraser uses 50\%  or more bit flips as the leakage signature for speculation. However, these bit flips can also result from non-leakage errors due to imperfect gates and measurements, especially when non-leakage errors are $10\times$ more likely than leakage errors. As illustrated in Figure~\ref{fig:patterns}(a), most patterns flagged by \eraser are probably due to non-leakage errors.

We evaluate \erasernspace's performance across eleven syndrome patterns, as depicted in Figure~\ref{fig:patterns}(a). We observe that \erasernspace's straightforward design fails to distinguish between different patterns effectively. Some patterns are much more likely to be caused by leakage than non-leakage errors. Therefore, we can classify these patterns into two groups, as, those requiring immediate mitigation and those that can be postponed without immediate intervention, allowing them to be addressed in the later QEC rounds.

Leakage and non-leakage errors can cause similar flips in the syndrome, complicating the leakage diagnosis. However, we can leverage the relative difference in the leakage and non-leakage event probabilities to help estimate the likelihood of error syndrome caused by leakage or non-leakage errors. Consider a syndrome with bit-flip pattern 0011 in the surface code, which could arise from multiple scenarios: \textit{(1)} Leakage Error on data qubit, causing 50\% bitflips due to malfunctioning CNOT gates, \textit{(2)} ``X" Error on data qubit, introduced after the second CNOT of the stabilizer circuit, \textit{(3)} Two independent bit flips during measurement, or 
\textit{(4)} Two independent errors on the last two qubits during the reset operation. Among these, leakage and ``X" errors on data qubit are first-order effects because they involve a single error event; the probability of these events is $c.P_{e}$, where $c$ is a constant and $P_{e}$ is the error probability. In contrast, scenarios involving two independent errors due to measurement or reset are second-order effects, requiring two concurrent error events with $c.P_{e}^2$ probability, making them substantially less likely, where leakage errors occur with probability $P_{leak}$ and non-leakage errors with $P_{e}$. We assume the leakage occurs with probability $\mathbf{P_{leak} = lr\times P_{e}}$, where $\mathbf{lr}$, \rev{leakage ratio}, is the scaling factor to represent intensity of leakage over other errors. 

\begin{figure}[h]
    \centering
    \includegraphics[width=\linewidth]{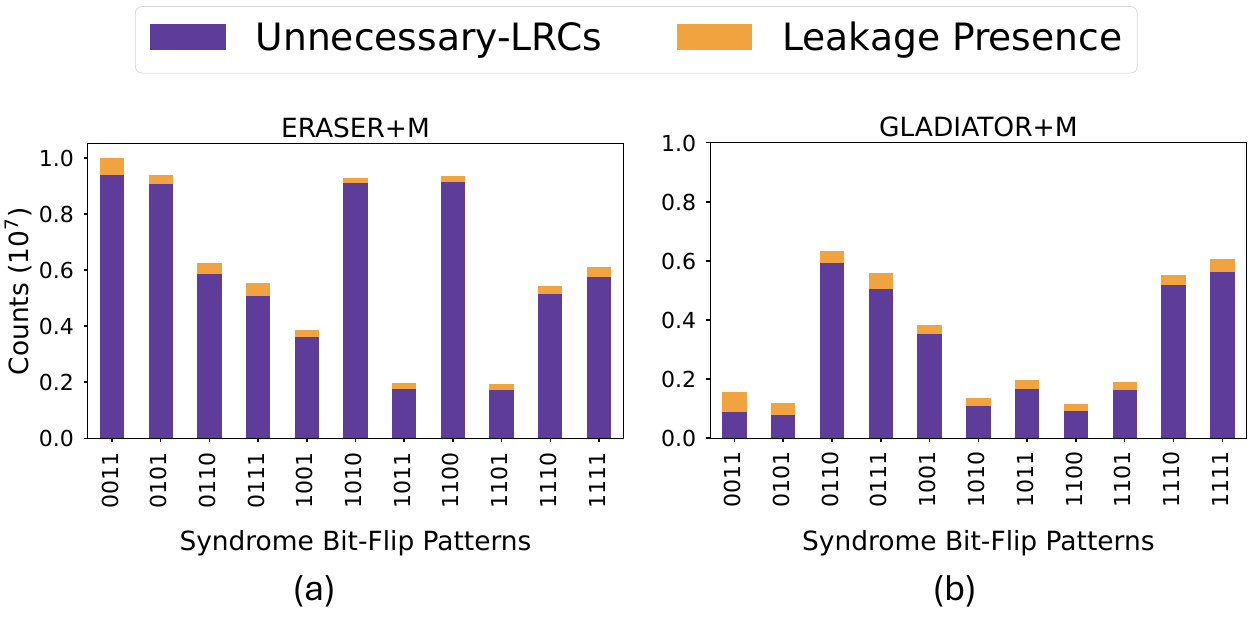}
    \caption{Unnecessary LRCs and leakage instances of (a) \eraserm (b) \oursm for 4-bit \eraser patterns }
    \label{fig:patterns}
    \Description{Counts of existing leakages and unnecessary LRCs applied for each of the speculation method}
\end{figure} 

\begin{figure*}[t]
    \centering
    
    \includegraphics[width=\linewidth]{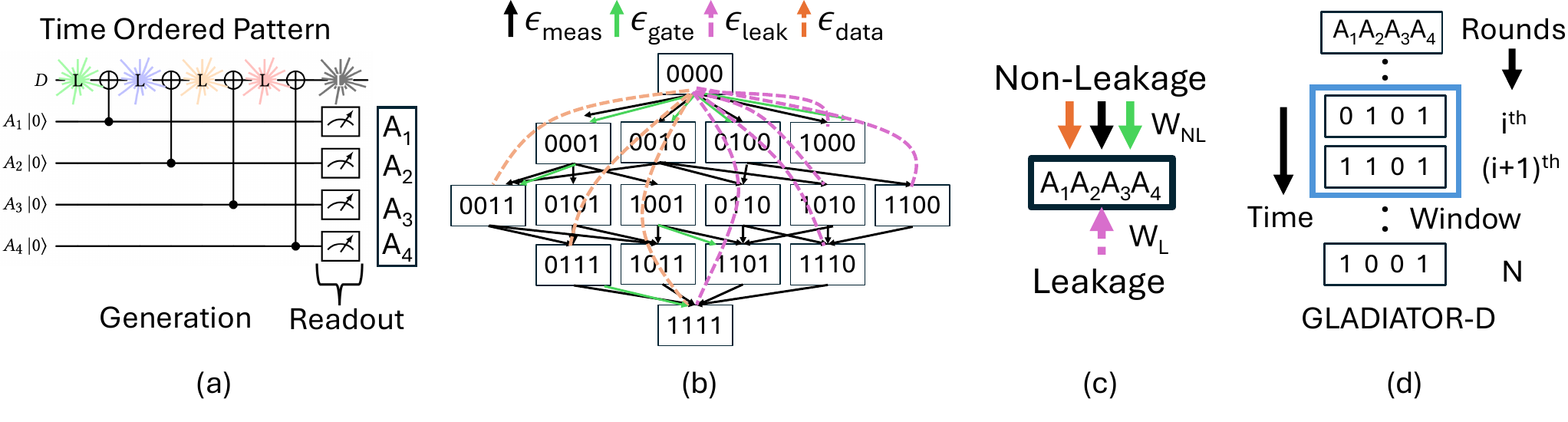}
    
    \caption{Overview of \ournspace. (a) Stabilizer circuit produces time-ordered syndrome patterns (b) A graphical model captures transitions between patterns based on error probabilities (c) Each node’s incoming and outgoing edge weights reflect likelihood of leakage over non-leakage for the pattern (d) \oursd uses a two-round windowed history for robust leakage speculation. }
    
    \label{fig:main}
    \Description{Overview of Surface code circuit for possibilities of leakage errors along with graph construction visualization for obtaining patterns using our method}
\end{figure*} 

In comparison, consider syndrome 0110. For the 0110 pattern, all non-leakage errors causes are second-order events, except where the ``X" Error on the data qubit generates the 1111 syndrome pattern and then followed by two measurement errors resulting in the 0110 pattern. On the other hand, if the data qubit leaks before the first CNOT or leaks before the second CNOT, we will observe the 0110 patterns with $\frac{1}{16}\times P_{leak}$ and $\frac{1}{8}\times P_{leak}$, respectively, which gives the combined probability of $0.02\%$ for $lr=0.1$ and $P_{e}=10^{-3}$ as shown in Figure~\ref{fig:main}(a), similarly, the probability that this syndrome is caused by non-leakage event is about $3.P_{e}^2+P_{e}^3=~10^{-6}$, Making leakage the primary cause to observe this pattern. We leverage this insight to determine the set of bit-flip patterns that most likely to indicate a leakage presence in the data qubit.

\subsection{GLADIATOR Design}\label{sec:design}
We formulate leakage diagnosis as a graph labeling problem.  In this model, each node represents a syndrome pattern. The edges between these nodes are weighted and directed, indicating how likely it is for one syndrome pattern to transform into another, a transformation driven by leakage or non-leakage errors. Our goal is to label the nodes of this graph as likely leakage and non-leakage patterns at post-qubit calibration and use the critical leakage patterns in the QEC controller to trigger LRCs judiciously at runtime.\footnote{Our approach parallels the standard decoders in inferring the most likely error by analyzing graph weights. However, most conventional decoders either lack awareness of leakage errors altogether or require multiple rounds of syndrome information before inferring the faults due to leakage. Our experiments highlights the cost of delaying detection of leakage faults, as leaked states persist across cycles, accumulate, and can propagate to neighboring qubits, ultimately causing logical errors. To prevent leakage accumulation, it is essential to quickly and accurately classify and detect leakage faults within two rounds from the occurrence of leakage.}

\textbf{Building Leakage and Non-Leakage Graphs.} The leakage graph shows the likelihood of leakage transforming one syndrome pattern into another. We build this leakage graph by starting with the base node, which is a syndrome pattern without any leakage, and we grow the graph by adding syndrome nodes that are transformed due to leakage error. This process is illustrated in Figure~\ref{fig:main}(b). We will use the base node 0000, a state without any error for simplicity.  During the generation and measurement of syndromes, any operation can introduce leakage.  Our experiments on IBM hardware show that leakage on control during CNOT can result in random bit-flips on target, which is a primary means of detecting data leakage using parity qubit measurement. Figure~\ref{fig:main}(a), shows the two qubit gates involved in surface code circuit with data qubit as D and the parity qubits in order as A1, A2, A3, and A4. As shown in Figure~\ref{fig:main}(a), leakage on data qubits just before the measurement cannot be detected until next round. However, if leakage occurs before \texttt{CNOT (D, A4)} (light-pink), we observe 50\%  bitflips in A4 due to malfunctioning CNOT, resulting in  0001 and 0000 syndromes. In comparison, if leakage is injected before \texttt{CNOT (D, A3)} (light-orange), we will observe 50\% bit flips in both A3 and A4, resulting in 0000, 0001, 0010, 0011 syndrome patterns. This process can be represented as a directed and weighted graph shown in Figure~\ref{fig:main}(b), where edge weights are a product of the base state probability (prior) and probability of leakage-driven transformation. The leakage events annotated in Figure~\ref{fig:main}(a) are color-coded to match the syndrome patterns discussed above.

In Figure~\ref{fig:main}(b), repeated nodes are merged by adding the edge weights corresponding to the pair of nodes. Figure~\ref {fig:main}(b) presents only a small number of edges for clarity. Next, we build leakage graphs starting with other base states that may have a non-leakage error, as shown in Figure~\ref{fig:main}(b). We merge all the individual graphs by summing edge weights for repeating source and target nodes.   

We use a similar methodology to build a non-leakage graph as the leakage graph. In the figure~\ref{fig:main}(b), we show how the base node "0000" can possibly be transformed into syndrome patterns due to first-order and second-order non-leakage errors. The non-leakage graph is reduced by merging repeating source and target node pairs and summing the edge weights, similar to the leakage graph's reduction. Using experimentally verified leakage (section~\ref{LeakageIBM}) and non-leakage error models, we can estimate how leakage errors are expressed in the syndrome measurements. 

\textbf{Graph Merging and Node Labeling.}
We merge the reduced leakage and non-leakage graphs; while merging, we merge the nodes of two graphs and keep all the non-leakage edges \rev{(colored $\epsilon_{gate}$ (green), $\epsilon_{measure}$ (black) and $\epsilon_{data}$(orange))  and leakage edges (colored $\epsilon_{leakage}$(pink))}. The final graph would be an all-to-all connected graph, with each node having both incoming and outgoing leakage and non-leakage edges, as shown in Figure~\ref{fig:main}(c). To label the node, we iterate over all nodes and sum the weights of all incoming non-leakage edges  to create a super edge $W_{L}$ such that $W_{L} = \sum w_{l}^{i}$as shown in Figure~\ref{fig:main}(c). On any node, if the leakage \rev{(pink)} super-node weight ($W_{L}$), which represents the cumulative likelihood that the syndrome is produced by leakage error, is greater than the weight of the non-leakage super-node ($W_{NL}$), i.e, cumulative likelihood that the syndrome is result of non-leakage error, by a threshold factor, we mark that node thereby the syndrome as leakage pattern for immediate mitigation in the next subsequent round. 

{\color{black} 

\ours operates in two stages: (1) Offline Stage: we analyze the target QEC circuit once, build an error-propagation graph, and weight its edges with calibration data (leakage rate, non-leakage noise, readout error). The result is a lookup table of syndrome patterns that strongly indicate leakage. (2) Online Stage, i.e, during QEC cycle, the syndrome bit pattern is used to query the table, and schedules an LRC only when the pattern is flagged as ``leakage".}

\subsection{GLADIATOR - Speculation Efficacy}
{\noindent{\textbf{Efficacy.}}} Figure~\ref{fig:patterns}(a-b) compares the LRCs inserted for observed syndrome patterns with leakage (golden bar) and without the leakage (purple bar). Compared to \erasernspace, both the \ours designs significantly improve the accuracy and thereby dramatically lower the number of LRCs used. \textit{Efficacy of} \ours \textit{is rooted in sophisticated leakage classification strategy. For instance, \eraser flags 11/16 4-bit syndrome patterns as leaked, including frequently occurring patterns such as ``0011'', most likely caused by non-leakage errors. In comparison,} \ournspace, \textit{flags 7/16 syndrome patterns excluding such frequently occurring patterns due to non-leakage errors. }

\rev{
\noindent\textbf{Tunable Features for New QEC Codes and Dynamic Errors.} 
\ours incorporate leakage propagation, error dynamics and stabilizer execution to build the graphical model as shown in Figure~\ref{fig:main}(b) to map the syndrome pattern transitions under both Pauli and leakage errors during the QEC code execution. Nodes represent syndrome patterns, while edges represent error combinations that transition one syndrome pattern to another. Non-leakage faults yield deterministic transitions, while leakage faults produce multiple possible outcomes incorporating leakage propagation dynamics (Section~\ref{LeakageIBM}). Each incoming edge encodes a distinct error path, with tunable weights, updated from calibrated error rates.\label{generalizability} 
}

\rev{
For newer codes with $n$-bit detector patterns, the transition graph is constructed by enumerating all possible errors before each gate in the stabilizer circuit, connecting base nodes to reachable detector flip patterns through annotated error edges similar to Figure~\ref{fig:main} (b). Recalibration updates only the edge weights, while preserving the underlying error paths in the graph structure, and allows \ours to adapt seamlessly to time-varying noise characteristics and hardware-specific error profiles. As the system evolves, the framework allows for incorporating additional noise sources as error transitions, added as edges to the graphical model during recalibration. \oursdm extends this framework using a two-round syndrome history as in Figure~\ref{fig:main}(d), to infer leakage and apply LRCs every round (except the first) in a sliding window fashion. 
}\label{Sec:G-D+M}

\rev{Large-scale quantum systems~\cite{AbuGhanem_2025, acharya2024quantumerrorcorrectionsurface, sycamore_leakage} exhibit considerable device variability, system drift and correlated noise, which can obscure the distinction between leakage-induced and standard Pauli errors. Unlike \eraser with fixed policy, \ours design with tunable graphical model delivers robust performance across a wide range of error profiles as discussed in section~\ref{error_profiles}, positioning \ours as a scalable and adaptable solution for leakage mitigation in diverse and dynamic QEC scenarios on quantum systems.
}

\subsection{Overheads of Deploying GLADIATOR}

Figure~\ref{fig:design_gladiator} illustrates the \ours microarchitecture. During each QEC cycle, syndrome data is processed by a parity adjacency generator to produce a syndrome pattern per data qubit. This pattern is evaluated by a sequence checker, which matches it against high-probability leakage signatures from \ournspace. If either a match is found or if the associated parity qubit is leaked, then the LRC scheduler triggers a leakage reduction circuit in the next round and notifies the QEC controller. Since common leakage patterns are stable across error rates, a single sequence checker can be efficiently shared across multiple data qubits.

\textbf{Data-Parity Adjacency Generator.} The design and functionality is illustrated in Figure~\ref{fig:design_gladiator}. Syndrome Data corresponds to measurement data from all the parity qubits. But the patterns deciphered by \ours pertains to individual data qubit. The leakage speculation aims to make predictions for each of the nine data qubits marked as $D_{i}$ in Figure~\ref{fig:design_gladiator}, by using the parity qubit $A_{i}$ measurement flips which interact with the data qubit. For example, the data qubit $D_{5}$ is connected to parity qubits ($A_{3}, A_{4}, A_{6}, A_{5}$). Upon measuring these parity qubits, we will get a four-bit string, say "1000", which is sent for sequence checking, which outputs a one-bit label, ``0" as leakage not present, and ``1" leakage present. 

\begin{figure}[b]
    \centering
    \includegraphics[width=\linewidth]{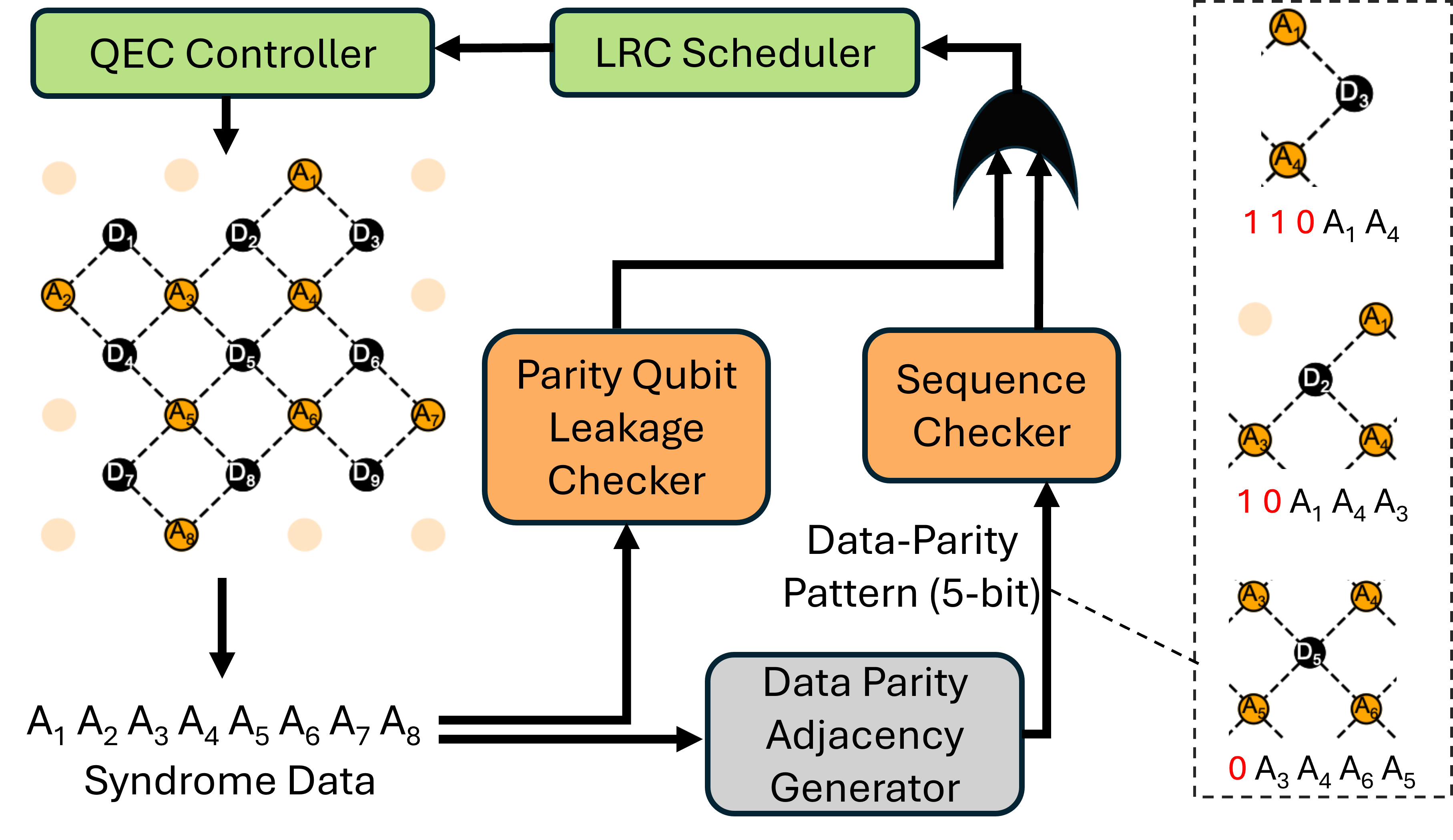}
    \caption{Overview of GLADIATOR's microarchitecture }
    \label{fig:design_gladiator}
    \Description{Design overview of our proposed method GLADIATOR}
\end{figure}

Note that not all data qubits have four adjacent parity qubits; corner qubits such as $D_{3}$ have just two parity qubits ($A_{1}, A_{4}$). To accommodate cases with two and three-parity qubits, we expand the sequence checker to match patterns for 2-bit, 3-bit, and 4-bit data-parity input to accommodate all possible syndrome patterns. For the patterns with 4-bit we prepend a index tag with ``0'', 3-bit with a ``10'' and 2-bit with ``110'' to make them all 5-bit patterns ($x_4x_3x_2x_1x_0$). This uniform representation simplifies processing while supporting all possible syndrome patterns. A network of multiplexers forms the core functionality of the Data-Parity Adjacency Generator.

\textbf{Sequence Checker.} 
\ours detects leakage by matching a 5-bit data-parity syndrome pattern against pre-defined Boolean templates using combinational logic. This enables fast, parallel evaluation with minimal hardware overhead. In contrast, \eraser relies on a hand-crafted finite state machine (FSM) to track syndrome history and trigger LRCs, resulting in high LUT usage that scales poorly with code distance.

\ours encodes all the common leakage patterns observed in surface codes as compact Boolean expressions. These patterns represent stable syndrome-flip signatures caused by persistent leakage. After  minimization, the pattern set reduces to:
\begin{small}
\begin{align*}
& (x_0 \wedge x_1 \wedge x_4) 
\vee (x_0 \wedge x_2 \wedge x_3) 
\vee (x_2 \wedge x_3 \wedge x_4) \\
& \vee (x_2 \wedge x_3 \wedge \neg x_1) 
\vee (x_2 \wedge x_4 \wedge \neg x_0 \wedge \neg x_1)
\end{align*}
\end{small}


This logic requires at most 10 LUTs per data qubit and can be evaluated within 1\,ns. To support code distance $d$, leakage detection must complete for all $d^2$ data qubits within 100\,ns—the approximate latency of four CNOTs on superconducting platforms like Google's~\cite{sycamore_leakage}. To meet this deadline, \ours replicates the sequence checker, therefore total LUTs per logical qubit:
\[
\text{LUTs}_{\text{total}} = 10 \times \left\lceil \frac{d^2}{100} \right\rceil.
\]
For example, at $d = 25$, \ours requires only 70 LUTs to detect leakage across all 625 data qubits in real time, demonstrating both scalability and low hardware cost. As shown in Table~\ref{tab:FPGA}, \ours achieves a \textbf{$17\times$ to $80\times$ reduction} in resource usage across \textbf{code distances 5 to 25}. We use the same Kintex UltraScale+ FPGA (xcku3p-ffvd900-3-e) as \textbf{\eraser}~\cite{ERASER}, and re-synthesize ERASER’s design to support larger distances for a fair comparison.

\begin{table}[t]
\begin{center}
\centering
\caption{\rev{LUTs Per Logical Qubit on Kintex UltraScale+ FPGA}}\label{tab:FPGA}
\setlength{\tabcolsep}{0.1cm}
\renewcommand{\arraystretch}{1.2}

\begin{tabular}{c c c c c c c}
        \hline
        \textbf{Method} & \textbf{d = 5} & \textbf{d = 9} & \textbf{d = 13} & \textbf{d = 17} & \textbf{d = 21} & \textbf{d = 25} \\
        \hline
        GLADIATOR        & 10   & 10   & 20   & 30   & 50   & 70   \\
        ERASER           & 177  & 633  & 1382 & 2434 & 3786 & 5393 \\
        Relative Reduction & 17.7x & 63.3x & 69.1x & 81.1x & 75.7x & 77.0x \\
        \hline
\end{tabular}
\end{center}
\end{table}

{\color{black}
LRC scheduling decisions must be made before the next round of error correction completes\footnote{Speculative scheduling and execution of LRCs makes synchronization~\cite{satvik2025synchronization} necessary}. Like \erasernspace, \ours meets this timing constraint, but does so using $17\times$ to $80\times$ fewer LUTs. This efficiency comes from using combinational logic to match 5-bit syndrome patterns against minimized Boolean templates, enabling fast, parallel detection with minimal overhead.}


\section{Generalizability of GLADIATOR}

\subsection{Leakage Detection Beyond Surface Codes}

Color codes~\cite{Takada_2024,Brown_2016, PhysRevResearch.6.033086}, QLDPC codes~\cite{QLDPC, kang2025quitsmodularqldpccode, LDPC_ibm} are emerging as a compelling candidate for quantum error correction, offering resource-efficient encoding of quantum information with significantly fewer qubits. The triangular 6.6.6 color code, as shown in Figure~\ref{fig:patterns_color_code}(b), constructed on a hexagonal lattice with a three-colorable structure, enables efficient fault-tolerant implementation due to its symmetrical stabilizer operations~\cite{PhysRevA.91.032330}. For instance, a code distance-7 color code 6.6.6 requires only 37 qubits compared to 97 qubits for a distance-7 surface code. This substantial reduction makes color codes particularly attractive for quantum systems with limited qubits.

Recent advancements by the Google QuantumAI team highlight the use of color codes in magic state cultivation~\cite{gidney2024magicstatecultivationgrowing}, showcasing their potential to reduce overhead in resource-intensive quantum algorithms. However, color codes introduce unique challenges for leakage detection due to their smaller syndrome bit patterns. In surface codes, each data qubit connects to four parity qubits, generating 4-bit syndrome patterns. In contrast, color codes often connect data qubits to fewer parity qubits, resulting in 3-bit patterns, with edge and corner qubits producing only 2-bit or 1-bit patterns, as illustrated in Figure~\ref{fig:patterns_color_code}(a). This limited information from single-round syndrome measurements reduces the accuracy of \ournspace, leading to unnecessary applications of leakage reduction circuits.

To address this limitation from syndrome measurements for leakage speculation, we introduce \oursdnspace, that incorporates temporal information by leveraging an additional round of measurements. By analyzing dependent syndrome patterns across time, as shown in Figure~\ref{fig:patterns_color_code}(d), \oursd enhances leakage speculation accuracy, reducing misclassifications and minimizing unnecessary LRCs. This integration of spatial and temporal data is particularly effective for complex QEC codes like color codes, where spatial information alone is insufficient. Leveraging temporal correlations, \oursd ensures precise leakage detection and robust mitigation, aiding in advancing fault-tolerant quantum computing.

\begin{figure}[t]
    \centering
    \includegraphics[width=\linewidth]{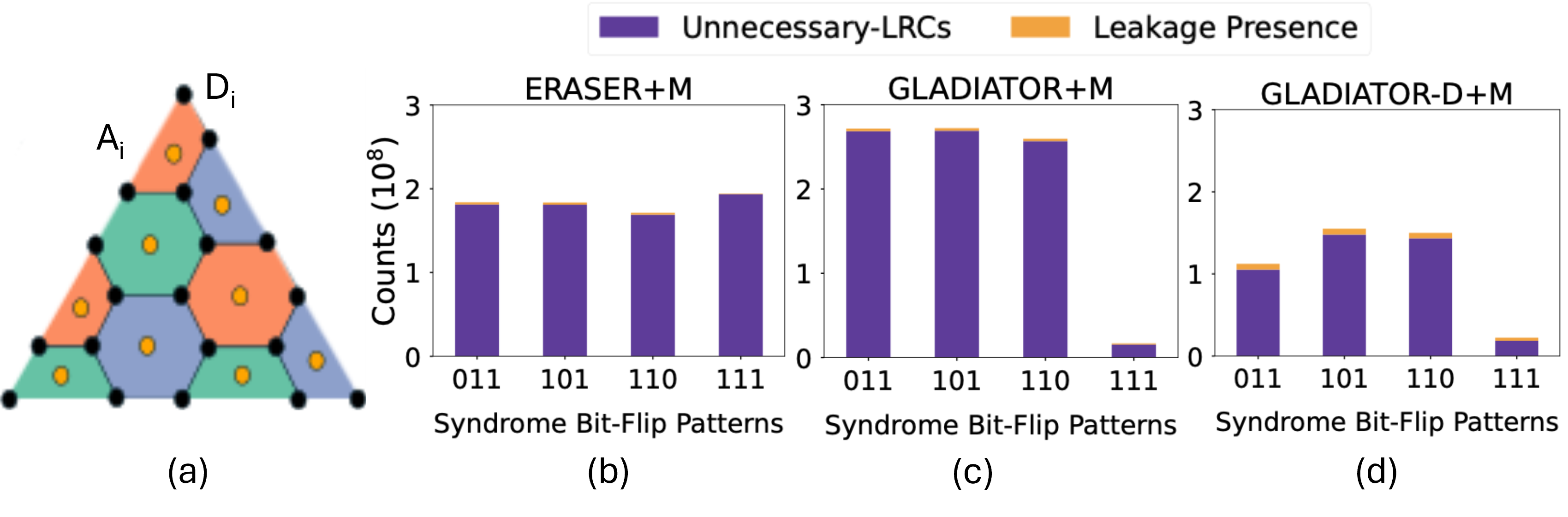}
    \caption{(a) Color Code (d=5), with data and parity qubits in black and orange. LRCs distribution (b) \eraserm (c) \oursm (d) \oursdm for 3-bit \eraser patterns }
    \label{fig:patterns_color_code}
    \Description{Understanding leakage instances and unnecessary LRCs for patterns flagged by ERASER for Color Code, along with overview of differences of distance-3 surface code and distance-7 color code}
\end{figure}

\subsection{Leveraging Syndrome Temporal Correlation}

Leakage speculation relies on syndrome patterns to detect leakage events, as leakage errors often persist across multiple QEC cycles. Deferring diagnosis by one additional QEC round can significantly improve detection accuracy by leveraging the evolution of consecutive syndrome patterns. Syndrome \( s_1 \) from round 1 influences the subsequent syndrome \( s_2 \) in round 2, depending on whether the error is caused by leakage or a non-leakage event. 

For instance, when the first-round syndrome pattern $s_{1}$, is ``0011,'' \eraser would typically insert an LRC, while \ours might not. This initial decision could be incorrect, as both leakage and non-leakage errors can produce the ``0011'' pattern. However, observing the pattern evolution can clarify the situation. If ``0011'' transitions to ``0011($s_{1}$) 1111($s_{2})$'' by next round, this pattern likely indicates a non-leakage \( X \)-error, such as one affecting the data qubit after the second CNOT in round 1. Conversely, if it evolves into ``0011 0101,'' it suggests leakage, as a leaked qubit causes random bit-flips that a single non-leakage error cannot explain.

Deferring leakage detection by one round significantly reduces false positives. \textit{For example, \oursdnspace, which uses two rounds (8-bit syndromes), flags 70 out of 256 patterns as leakage, compared to 121 out of 256 flagged by \eraser.}

Deferring leakage speculation to gather more information is particularly effective for color codes, which have limited syndrome information due to smaller patterns: 2-bit for edge qubits, 1-bit for corner qubits, and 3-bit for the rest. This limitation causes \eraser to over-apply Leakage Reduction Circuits (LRCs), as shown in Figure~\ref{fig:patterns_color_code}(b). For example, out of all 3-bit patterns, \eraser flags 4 out of 8 patterns as leakage, while \ours flags 3, offering only slight improvement shown in Figure~\ref{fig:patterns_color_code}(c). By leveraging temporal correlations, \oursd captures leakage propagation effects more effectively. It flags 11 out of 64 patterns compared to 16 flagged by \eraser, as shown in Figure~\ref{fig:patterns_color_code}(d). For instance, in the 3-bit pattern "010," observing its evolution to "010 111" suggests a non-leakage error, while "010 110" indicates leakage. This approach enhances the resilience of color codes by reducing unnecessary LRCs.

\textbf{Design modifications.} 
To support \oursdnspace, which leverages multi-round syndrome history, we extend the design to enable deferred leakage detection. This version builds leakage and non-leakage graphs from two consecutive QEC rounds, expanding the input to the sequence checker from 5-bit to 10-bit patterns. These extended patterns, generated by the Data Parity Adjacency Generator, improves the leakage diagnosis to mitigate leakage while increasing LUT overhead by at most 4$\times$.

\section{Methodology}
\label{sec:meth}

{\textbf{Circuit Noise Model.}} We assumes a physical error rate of $p$, accounting for several key error sources. Data qubit depolarization occurs at the start of each QEC round with a probability $p$. We apply readout errors during qubit measurement, while gate operation faults apply depolarizing errors to qubit operands after each $CNOT$ or $H$ gate, both with a probability $p$. Initialization errors occur during qubit resets, also modeled with a probability $p$.

{\textbf{Leakage Errors and Propagation.}}\label{sec:error_rates} Leakage errors, primarily caused by two-qubit gates such as CNOT operations, include environment driven leakage injected into data qubits at the start of each QEC round with a probability $p_{l}$, and leakage from gate operations occurring with the same probability. \textit{Leakage propagation} is modeled with a 10\% probability, transferring leakage to the target qubit during CNOT operations from control qubits. In the remaining 90\% of cases, a random Pauli error $(I,X,Y,Z)$ is applied to the target qubit to model random bit-flip errors due to \textit{gate-operation induced leakage}. Additionally, \textit{environment-induced leakage} is modeled by injecting leakage into data qubits at the beginning of each QEC round with a probability $p_{l}$. Our IBM hardware characterization in Section~\ref{LeakageIBM}, confirms similar leakage transport and bit-flip effects.

{\textbf{Modeling Error Profiles.}}\label{leakage_ratio} We extend the ERASER framework by introducing the leakage ratio $(lr=\frac{p_{l}}{p})$ to quantify the intensity of leakage errors relative to non-leakage errors. Similarly we use $mlr$ for Multi-Level Readout to model readout errors for leaked states as $mlr\times p$. For evaluations, we assume $lr=0.1$ and $mlr=10$, representing a $10\times$ higher readout error for $\ket{2}$.\footnote{This is a conservative assumption. Several experimental works~\cite{QutritReadout_autoencoders, mude2025efficientscalablearchitecturesmultilevel} show MLR inaccuracies to be at most 2x worse compared to two-level readout.}  We test across $lr = {0.01, 0.1, 1}$ to ensure adaptability across diverse noise conditions.

{\textbf{Interplay of Leakage with other Noise Sources.}}\label{sec:interplay_noises} We simulate stabilizer circuits with errors introduced at each step of the QEC cycle. Initialization and leakage errors occur at the start of each round, while CZ gates introduce both depolarizing and leakage errors. Measurement errors include those arising from multi-level measurement in the presence of leakage. This comprehensive modeling captures the interaction of leakage with other noise sources, evaluating their combined impact on QEC performance.

{\textbf{Simulation Framework.}} We modified the \eraser artifact for our evaluation. We also use Google's Stim~\cite{STIM} simulator to simulate leakage errors through the Tableau simulator.

\textbf{Quantum Hardware.}{\color{black} We conducted leakage characterization on 7-qubit IBM machines (Lagos, Jakarta, Perth) using Qiskit and Qiskit Pulse~\cite{Qiskit}. Each experiment used 10,000 shots. Leakage was induced by applying an $X$ gate followed by a calibrated $\ket{1} \rightarrow \ket{2}$ pulse, following IBM’s official guide~\cite{IBMGuide}. These experiments were performed before IBM retired pulse-level access in September 2024~\cite{IBMPulseRetire}. Despite this, the observed leakage effects are consistent with prior studies~\cite{sycamore_leakage, NC_critical_faults, SC_leak_effect_QEC, coping_qubit_leakage}. 
}


\textbf{Scaling Simulations 
using Leakage Sampling.} 
As leakage spreads and population grows over time, evaluating performance for larger QEC cycles becomes critical. However, \erasernspace’s methodology is limited to 10 QEC cycles due to the computational cost of capturing meaningful leakage events. For  $p_{l}=10^{-4}$ and $p=10^{-3}$ and a code distance $d=7$, only 0.5\% of runs begin with at least one qubit in a leaked state, making it inefficient for long running evaluations. To address this, we introduce leakage sampling, where simulations start with at least one leaked data qubit to focus on scenarios where leakage is actively present. 

This method of leakage sampling allows us to extend simulations to 100$\times$ more QEC cycles while significantly reducing computational overhead. This method ensures that the performance of speculation policies is evaluated under relevant conditions, particularly when leakage is present. We evaluate our policies over 100 QEC cycles using leakage sampling, with \eraser as the baseline.


\section{Evaluations}

To assess the efficacy of our design, we utilize - \textit{(1) Data Leakage Population Ratio (DLP):} The average fraction of leaked data qubits per round, computed across all simulation shots. This metric quantifies sustained leakage over $n$ QEC rounds, assessing the effectiveness of leakage mitigation. \textit{(2) Total LRC Usage Rate:} The total number of LRCs applied per round across all simulation runs. \textit{(3) False Negatives and False Positives:} The number of undetected leakages and unnecessary LRC applications.\textit{(4) Logical Error Rate (LER):} The probability of a logical error after decoding the syndromes.

\begin{figure}[b]
    \centering
    \includegraphics[width=\linewidth]{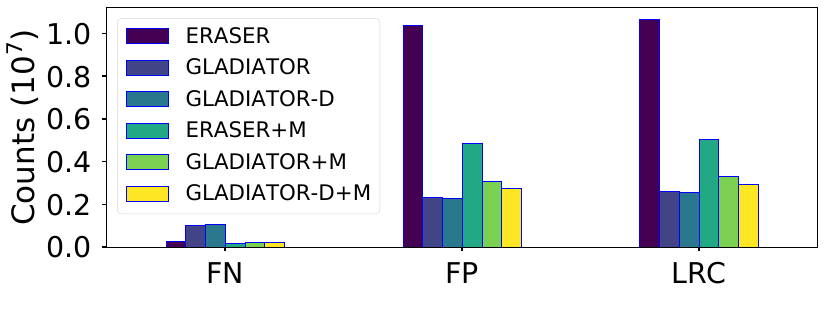}
    \caption{False Negative (FN), False Positive (FP), Leakage Reduction Circuit (LRC) counts with the leakage error as $p_{leak}$ = $10^{-4}$, gate error p=$10^{-3}$ for surface code (d=7)}
    \label{fig:LRCs}
    \Description{Comparison of False Negative(FN), False Positive(FP), Leakage Reduction Circuits (LRC) counts for different methods}
\end{figure}

\subsection{Speculation Accuracy and LRC usage}

The efficacy of leakage speculation hinges on efficient mitigation of leakage through optimal LRC insertion. False Positives (FPs) introduce unnecessary physical errors and increase execution time, while False Negatives (FNs) represent undetected leakages that allow leakage to spread, leading to accumulation. Multi-Level Readout (MLR) lowers FNs by detecting leakage transport, preventing leakage accumulation. MLR errors can introduce FPs unrelated to syndrome patterns, necessitating better MLR accuracy. Together, lower FPs, FNs, and LRC usage directly translate to improved error rates and better overall system performance.

\ours and \oursd significantly reduces the false positive rate, i.e, the number of un-necessary LRCs inserted compared to the baseline \erasernspace. Figure~\ref{fig:LRCs} shows the False Negative (FN) rate, False Positive (FP) rate, and total LRCs inserted by different policies for leakage ratio $lr=0.1$, code distance 7 and error probability $10^{-3}$, similar to \erasernspace. Compared to \erasermnspace, \oursm reduces the false positive rate by $1.56\times$, and \oursdm by $1.76\times$ with an increase in false negatives by $1.16\times$, $1.22\times$, leading to a reduction in number of LRCs inserted by $1.53\times$ and $1.71\times$ respectively. 

\begin{figure}[ht]
    \centering
    \includegraphics[width=\linewidth]{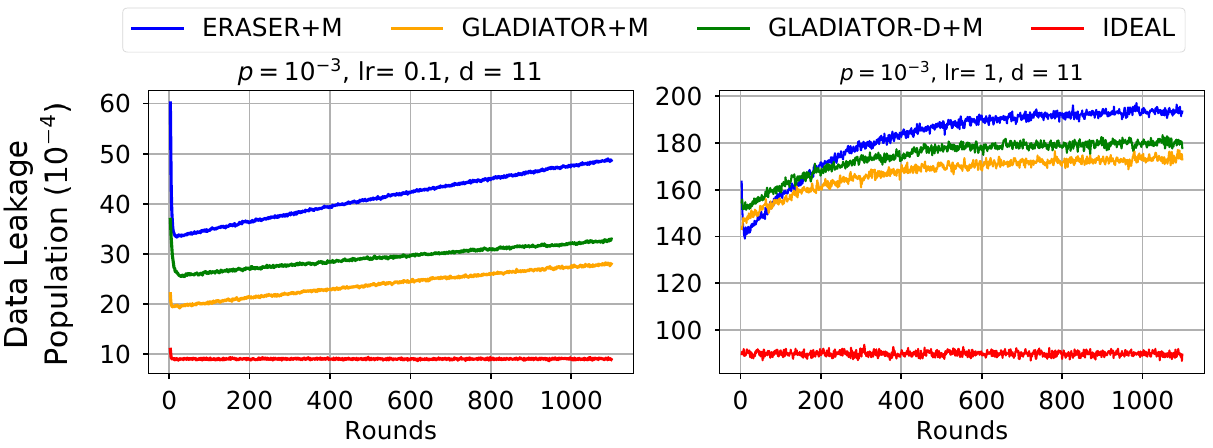}
    \caption{Data leakage (lower is better) for code distance $d$, leakage ratio $lr$ and run for $100 d$ rounds and $10^{5}$ repetitions}
    \label{fig:DLP_error_profile}
    \Description{Comparison of data leakage population for GLADIATOR, GLADIATOR+M vs ERASER+M speculation methods}
\end{figure}

\begin{figure}[ht]
    \centering
\includegraphics[width=\linewidth]{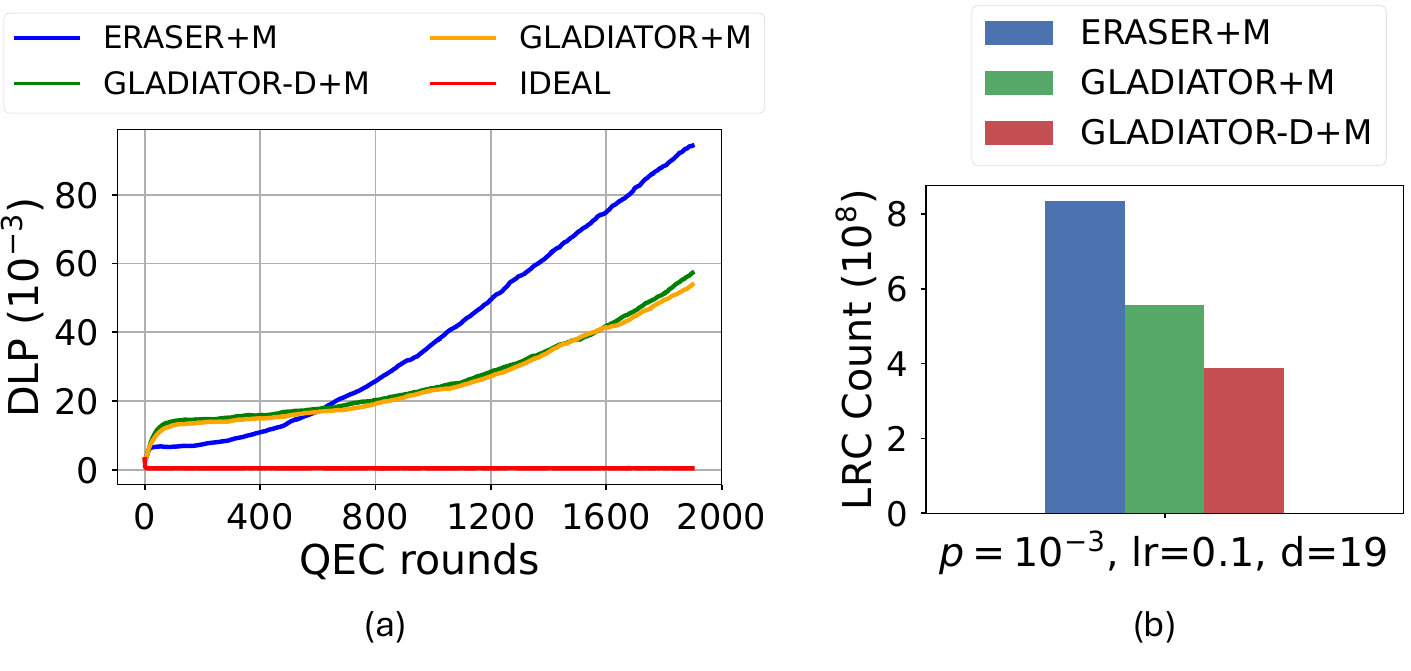}
    \caption{Comparison of (a) Data Leakage Population (b)  LRC usage for color code with d=19 run for 100 QEC cycles}
    \label{fig:color_code_d19_dlp}
    \Description{Data leakage population for color code with code distance 19}
\end{figure}

\subsection{Impact on Data Leakage Population}

Short-term simulations over 10 QEC cycles as shown in \erasermnspace, fail to reveal the leakage accumulation over extended rounds. To better understand this accumulation, we evaluate the performance of \oursm and \oursdm over 100 QEC cycles. For a higher leakage intensity $(lr=1)$, we observe a crossover point between \eraserm and the proposed policies, \oursm and \oursdmnspace, occurring between 100 and 200 rounds, for \textbf{surface codes}, as shown in Figure~\ref{fig:DLP_error_profile}. Additionally, Figure~\ref{fig:color_code_d19_dlp}(a) also shows the increasing gap between \oursm and \erasermnspace, with increasing QEC rounds for \textbf{color codes}. Leakage growth differs over time, with a slower leakage accumulation rate for \oursdm and \oursm compared to \erasermnspace.

Specifically, for code distance $d=11$ and $lr=0.1$, the data leakage population for \oursdm and \oursm is $1.47\times$ and $1.73\times$ lower than \erasermnspace, respectively, after $100\times d$ QEC rounds. This demonstrates superior leakage mitigation due to better speculation. The slower rate of leakage accumulation for \oursdm ($1.14\times$ lower than \oursmnspace) indicates the advantages of deferred speculation for sustaining extended QEC runs of thousands of rounds. While in surface codes \oursm reduces leakage more overall, the finer control offered by \oursdm makes it a be alternative for mitigating leakage growth over extended QEC cycles especially for color codes.

\begin{figure}[b]
    \centering
\includegraphics[width=\linewidth]{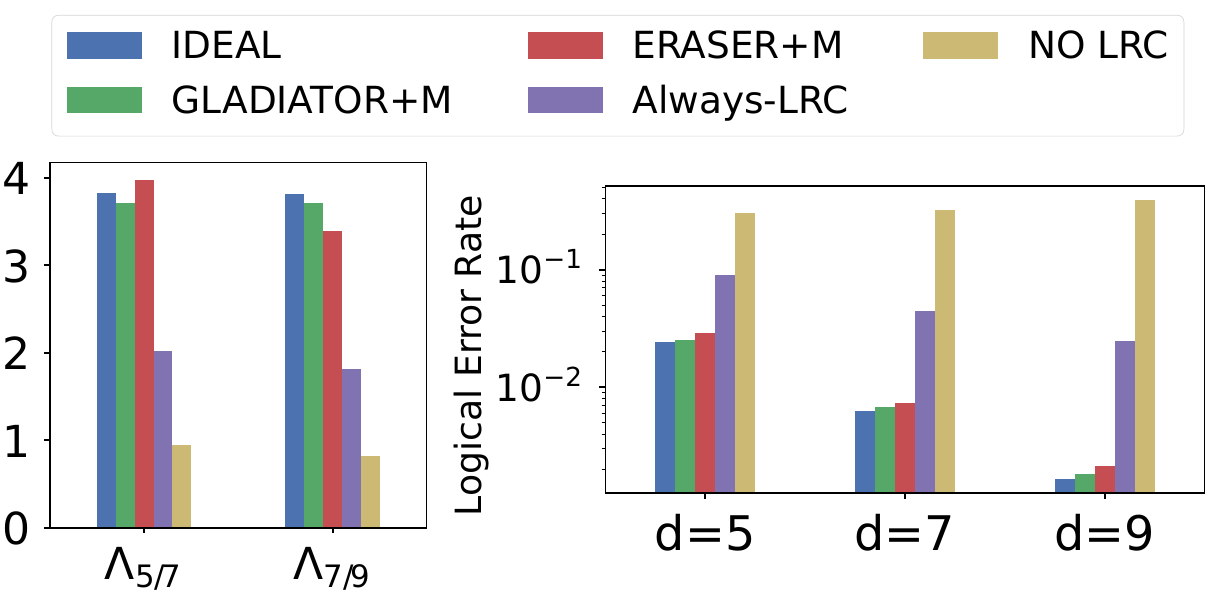}

    \caption{ \rev{Logical Error Rate (LER) for increasing code distances d =5,7,9 run for $10d$ cycles with $p=10^{-3}$ and $lr=0.1$}}
    \label{fig:LER_comp}
    \Description{Logical Error Rate (LER) of code distances d = 7,9,11 run for 10 QEC cycles with $p=10^{-3}$}
\end{figure}

\subsection{Impact on Logical Error Rate} \label{Sec:LER_evals}

Figure~\ref{fig:LER_comp} compares the logical error rate (LER) across three leakage mitigation strategies—\oursm, \eraserm, and {\textsc{Always-LRC}}—across multiple surface code distances. We observe that closed-loop approaches significantly outperform open-loop strategies, with \oursm achieving up to an order of magnitude lower LER than {\textsc{Always-LRC}} at comparable code distances.

We evaluate the suppression factor $\Lambda_{d/(d+2)} = \frac{\epsilon_{d+2}}{\epsilon_{d}}$, to quantify scalability by capturing the rate of logical error reduction as the code distance increases. For distances $d=\{5,7,9\}$, \oursm achieves an average $\Lambda$ of approximately 3.71, compared to 3.38 for \eraserm—indicating better error suppression with increasing code size. This improvement is closely tied to better leakage control. As leakage error rate decreases, \oursm adaptively reduces unnecessary LRC applications, resulting in nearly $2\times$ lower residual leakage population and speculation inaccuracy (see Table~\ref{tab:leakage_specinacc_corrected}) translating to improved LER for $p={10^{-4}}$ as shown in Figure~\ref{fig:operational_error}(a). 

\begin{table}[ht]
\begin{center}
\small
\caption{{Total Leakage Equilibrium across Leakage Ratios and Speculation Inaccuracy across Physical Error Rates ($d=11$)}}
\setlength{\tabcolsep}{0.15cm}
\renewcommand{\arraystretch}{1.3}
\begin{tabular}{cccccc}
\hline
\multirow{2}{*}{\textbf{Method}} & \multicolumn{3}{c}{\textbf{Leakage Equilibrium}} & \multicolumn{2}{c}{\textbf{Spec. Inaccuracy}} \\
 & \textbf{lr = 0.01} & \textbf{lr = 0.1} & \textbf{lr = 1.0} & \textbf{p = $10^{-3}$} & \textbf{p = $10^{-4}$} \\
\hline
GLADIATOR+M & 0.00148 & 0.00386 & 0.01538 & 0.01335 & 0.00238 \\
ERASER+M    & 0.00276 & 0.00660 & 0.01782 & 0.02606 & 0.00736 \\
\hline
\end{tabular}
\label{tab:leakage_specinacc_corrected}
\end{center}
\end{table}

\rev{To highlight the cost of unmitigated leakage, we include a NO-LRC baseline in Figure~\ref{fig:LER_comp}. Unlike other schemes, its LER increases with code distance, almost $200\times$ \ours at $d=9$, to demonstrate leakage accumulation without LRCs degrading QEC. 
} \label{Sec:NO-LRC}

\subsection{Generalizes to Color Codes and LDPC Codes} \label{sec:QLDP_Data}

Table~\ref{tab:Generalizability} shows that \ours outperforms ERASER across a range of codes by reducing the number of unnecessary LRCs, lowering residual data-leakage population (DLP), and enabling faster QEC execution. While both methods perform well on the surface code, \ours still achieves $1.7\times$ fewer LRCs, $1.67\times$ lower DLP, and $1.69\times$ faster QEC execution. These benefits are even more pronounced on Hypergraph Product (HGP) codes, \ours reduces LRC count and QEC execution time by nearly $4\times$, and lowers DLP by $1.88\times$. Fewer LRCs not only reduce circuit depth but also shorten the QEC cycle, as LRCs are typically inserted after entangling gates and extend the duration of quantum error correction rounds. 

To quantify QEC performance, we measure the cycle time as the total QEC execution time normalized by the number of rounds and shots. We further compute the average LRC count per round per shot and convert this into an estimated latency overhead, assuming SWAP-based LRC implementations. This allows us to approximate the time cost directly attributable to leakage mitigation. We then compare the normalized QEC cycle times for \ours and ERASER across different codes. This analysis reveals the overhead reduction achieved by \ours through more selective LRC application, resulting in faster and more efficient QEC execution.

Unlike \erasernspace, which is tailored to surface-code symmetry, \ours generalizes across surface code, color code, Hypergraph Product (HGP) code, and Balanced Product Cyclic (BPC) code using only syndrome history. These results highlight \ournspace’s effectiveness and broad applicability to heterogeneous quantum error correction codes, the direction pursued by Google, IBM in their push towards large-scale, scalable quantum computing.
\begin{table}[ht]
\begin{center}
\small 
\centering
\caption{\textcolor{black}{Reduction Factors for \ours over \eraser}}\label{tab:Generalizability}
\setlength{\tabcolsep}{0.05cm}
\renewcommand{\arraystretch}{1}

\rowcolors{2}{lightgray}{lightgray}  
\begin{tabular}{>{\color{black}}c >{\color{black}}c >{\color{black}}c >{\color{black}}c >{\color{black}}c}
        \hline
        \textbf{Metric / Code} & \textbf{Surface Code} & \textbf{Color Code} & \textbf{HGP Code}  & \textbf{BPC Code} \\
        \hline
        LRCs        & 1.71x   & 1.5x  & 3.86x  &   1.51x \\
        DLP         & 1.67x   & 1.54x & 1.88x  &   1.02x \\
        {\color{black}\textbf{ QEC Cycle Time}}     & 1.69x   & 1.5x  & 3.83x  &   1.505x \\
        \hline
\end{tabular}
\end{center}
\end{table}

\begin{figure}[ht]
    \centering
    \includegraphics[width=\linewidth]{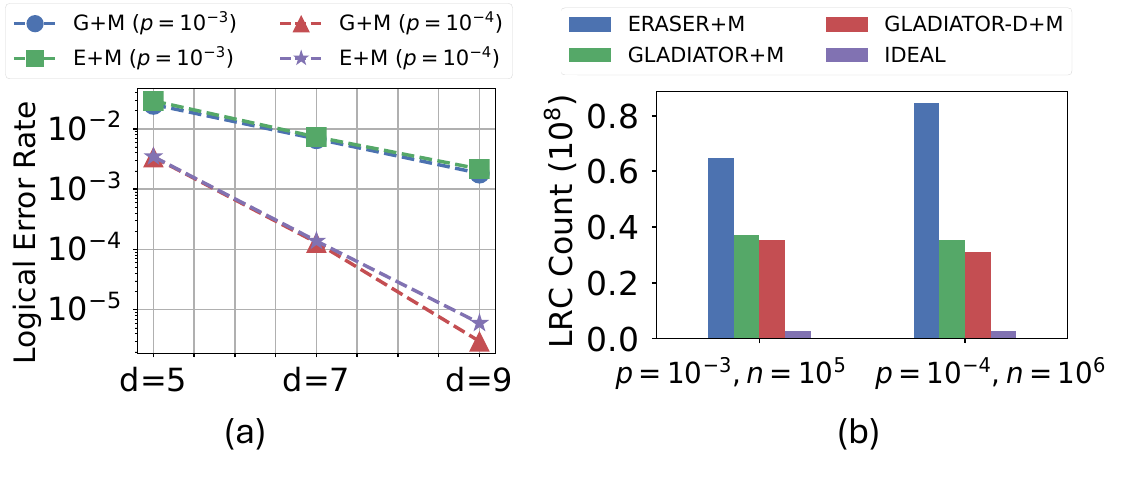}
    \caption{Comparison of (a) Logical error rate (b) LRCs usage for error probability $p=10^{-3}$ and $p=10^{-4}$ run for $n$ shots}
    \label{fig:operational_error}
    \Description{Impact on data leakage population due to operational error p}
\end{figure}

\subsection{Sensitivity Study}\label{error_profiles}

\noindent{\textbf{Sensitivity to Error Rate:}} Leakage occurrence directly correlates with LRC usage and LER, with higher $p_{leak}$ leading to increased LRC counts per shot and sustained leakage.
Similarly, as operational error probability $(p)$ decreases, both LER (Figure~\ref{fig:operational_error}(a)) and LRC usage (Figure~\ref{fig:operational_error}(b)) decreases. Among the evaluated methods for LRC counts, \oursdm shows superior adaptability by leveraging additional rounds for effective leakage management. Table~\ref{tab:leakage_specinacc_corrected} shows the leakage population at the equilibrium state is lower for \oursm compared to \eraserm for different $lr$.

\begin{figure}[t]
    \centering
\includegraphics[width=\linewidth]{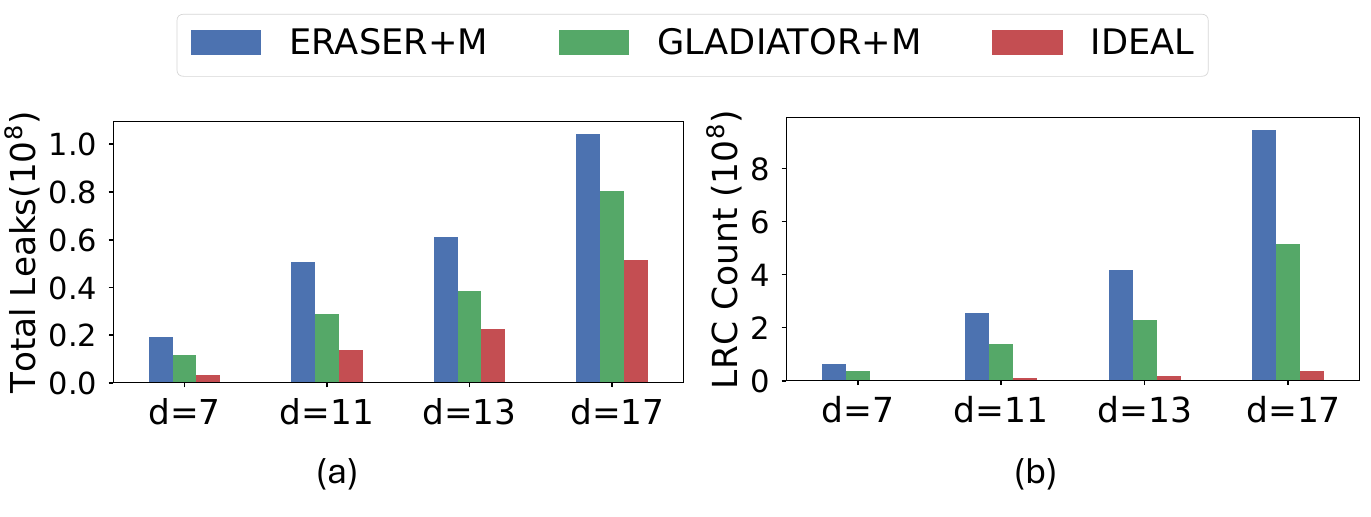}
    \caption{  For code distances d = 7,11,13,17 run for $100d$ cycles the plots show a) Total leakages and b) Total LRCs utilized}
    \label{fig:varying_d_mDLP}
    \Description{Total leakage and Total LRCs employed for surface codes with code distances 7,11,13,17}
\end{figure}

\noindent{\textbf{Sensitivity to Code Distance:}} While error rates generally decrease with increasing code distance $d$, the total leakage does not follow the same trend. Figure~\ref{fig:varying_d_mDLP}(a) illustrates that even under an ideal policy, assuming perfect speculation, leakage is still introduced by LRCs and the syndrome generation circuit due to the quadratic growth in the number of qubits and gates required as the code distance scales. Our evaluations show that qubit leakage becomes increasingly pronounced as the code distance grows. This increase in total leakage directly translates to a higher count of LRCs needed for leakage mitigation. LRC usage gap between \oursm and \eraserm widens with increasing code distance, demonstrating the efficiency and scalability of our leakage mitigation strategy. 


\noindent{\textbf{Impact on Execution Depth.}}\label{execution_depth} {\sc always-lrc} applies 60 LRCs per round, increasing execution depth by $20\%$ for  d=11. \ours ~applies only 1.22 LRCs per round ($\approx50\times$ fewer), resulting in only $0.4\%$ increase in execution depth compare to no LRC. The reduction factor in LRCs per round grows from $\approx40\times$  at $d=7$ to $\approx50\times$  at $d=17$, showing \ours~ is more effective at larger code distances.

{\color{black}
\subsection{Speculating Leakage Mobility}

On real quantum hardware, both how often qubits leak and how easily that leakage spreads (leakage mobility) can vary. This plays a crucial role in how well different mitigation strategies work. For example, walking surface codes use extra operations to move leakage from data qubits to ancilla qubits, where it can be detected and reset. But this only works well when leakage mobility is low and the leakage stays put unless intentionally moved \cite{camps2024leakagemobilitysuperconductingqubits}.

To adapt to different hardware behaviors, we introduce a practical way to estimate leakage mobility. Our approach combines \ournspace’s speculative detection, which flags likely leaked data qubits, with multi-level readout (MLR) signals that reveal leakage on ancilla qubits. By measuring how often an ancilla qubit is leaked when its neighboring data qubit is flagged, we estimate the leakage mobility. Based on prior work~\cite{camps2024leakagemobilitysuperconductingqubits}, we use a 5\% threshold: if the conditional probability is below 5\%, we classify the system as having low mobility; if it’s above, we consider it high mobility.

\begin{table}[ht]
\centering
\caption{Leakage Mobility Classification via \ours}
\setlength{\tabcolsep}{4pt}
\renewcommand{\arraystretch}{1.2}
\begin{tabular}{l | c c c c c}
\hline
\textbf{Mobility (\%)} & 1.0 & 2.5 & 5.0 & 6.0 & 9.0 \\
\hline
True Regime & Low & Low & High & High & High \\
Accuracy (\%) & 100 & 100 & 50 & 100 & 100 \\
\hline
\end{tabular}
\label{tab:mobility_classification}
\end{table}

Our method accurately classifies leakage mobility regimes, helping determine whether the system is best suited for open-loop or closed-loop mitigation. Low mobility supports simpler strategies like Staggered Always-LRC or walking codes to prevent leakage buildup. High mobility, on the other hand, favors dynamic, feedback-driven approaches like \ournspace.}

\section{Conclusion}
{\color{black}
Leakage errors pose a fundamental challenge to fault-tolerant quantum computing by corrupting syndrome extraction. While prior mitigation strategies like \eraser apply leakage reduction circuits (LRCs) speculatively using a fixed heuristic, they suffer from high false positive rates, leading to unnecessary LRC insertions that degrade performance and reliability.

This paper introduces \ournspace, a principled, model-driven framework that speculates leakage events by analyzing syndrome patterns through an offline-constructed, code-aware error graph informed by device calibration data. \ours approach generalizes across QEC codes—surface, color, and beyond—adapting to hardware-specific noise profiles without relying on hardcoded rules or code-specific assumptions. Through accurate leakage speculation, \ours significantly reduces false positives and unnecessary LRCs, achieving up to $3\times$ fewer LRCs, 16\% lower logical error rates, and $1.7 - 3.9\times$ shorter QEC cycles, resulting in significant speedups in application runtimes compared to \eraser~\cite{ERASER}.}

\begin{acks}
The authors would like to thank Benjamin Lienhard, Satvik Maurya, and Ayushi Dubal for their feedback on previous versions of this paper. The authors also thank Professor Guri Sohi for the help with the computing infrastructure. This research was supported by an NSF CAREER Award $\#2340267$, and the Office of the Vice Chancellor for Research (OVCR) at the University of Wisconsin-Madison, with the support of Wisconsin Alumni Research Foundation (WARF).
\end{acks}




\bibliographystyle{ACM-Reference-Format}
\bibliography{References}

%
%
%
%
%




\appendix
\section{Artifact Appendix}

\subsection{Abstract}
This artifact contains the code to verify the key results of this paper.
Specifically, this artifact will generate and plot data for Figures 4(b),
9, 10, 11, 12, 13 and 14. Additionally, the artifact contains code to
verify the results of Table 2. 

\subsection{Artifact check-list (meta-information)}

{\small
\begin{itemize}
  \item {\bf Algorithm: GLADIATOR, a leakage-detection algorithm.}
  \item {\bf Compilation: GCC}
  \item {\bf Binary: leakage}
  \item {\bf Metrics: LRC Usage, Logical Error Rate (LER), Data Leakage Population (DLP), False Positivies and False Negatives}
  \item {\bf Output: Data files and figures.}
  \item {\bf How much disk space required (approximately)?: Not more than 5GB}
  \item {\bf How much time is needed to prepare workflow (approximately)?: About a minute}
  \item {\bf How much time is needed to complete experiments (approximately)?: 15-20hrs}
  \item {\bf Publicly available?: Yes}
  \item {\bf Code licenses (if publicly available)?: MIT License}
  \item {\bf Archived (provide DOI)?:  https://doi.org/10.5281/zenodo.16735148}
\end{itemize}
}

\subsection{Description}

\subsubsection{How to access}
The artifact for this work is available on Zenodo at  https://doi.org/10.5281/zenodo.16735148


\subsubsection{Software dependencies}
The code is built by modifying the artifact from ERASER\cite{ERASER} using CMake v3.20.3. The compiler g++-12 and g++-13 are used for the evaluations, and to parallelize the experiments on computing clusters we use OpenMPI v4.x.x. CMake is used to package all other de-
pendencies within the code.

Additionally for plotting the figures, we provide a python notebook, tested using Python v3.12, and the following packages are dependencies: matplotlib v3.6.1, numpy v1.23.4, scipy v1.9.2, and pandas v2.3.0, though it should work fine with newer versions.



\subsection{Installation}

For creating the data for Figures 4(b), 9, 10, 11, 12, 13 and 14, we encourage to use a build directory `build' to avoid any issues. The executable leakage can be
generated as follows,

\begin{highlightitembox}
\begin{verbatim}
$ cd build
$ cmake .. -DCMAKE_BUILD_TYPE=Release
$ make -j8   
\end{verbatim}
\end{highlightitembox}


\subsection{Experiment workflow}
We explain how to generate the data for the main insights and results of this work, i.e, Figures 4(b), and from Figure 9 till 14. For each of the Figures, we provide {\texttt{AE\_Figure\_<\#>.sh}} to generate the data for the corresponding figures. Additionaly, we also provide {\texttt{ AE\_all\_data\_for\_plots.sh}} to generate all the data needed to plot the results, as follows,

\begin{highlightitembox}
\begin{verbatim}
$ cd leakage
$ ./AE_all_data_for_plots.sh <PROC> 
\end{verbatim}
\end{highlightitembox}
where PROC is the number of processors used by OpenMPI to parallelize the experiments. This will create data for plotting in the {\texttt{ leakage/AE\_results/ folder}}.  

Our evaluations are performed at $p = {10^{-3}}$ and at $p = {10^{-4}}$. We note that due to the lower logical error rate of $p = {10^{-4}}$, evaluations will take much longer due to requiring significantly more shots.
We recommend 100k shots for $p = {10^{-3}}$ and 1M shots for $p = {10^{-4}}$, as 100k is insufficient to obtain complete results for LER at $p = {10^{-4}}$.

We recommend using a cluster with atleast 32 cores with sufficient memory, as the memory requirement to run experiments with larger distance codes is significant and may need many cores to complete in
time. For reference, our evaluations for Figure 14 took total of 8 hours running on a cluster with 64 cores. The memory requirement arises from the need for storing measurement data for each experiment to run LER evaluations. As the code distance (d) increases to 17 and run for 1700 (100d) rounds for 1M shots, the memory requirement increases to store increase over 20GB. 

For Figure 14, as run experiments for code distance (d) till 17, and for 100d rounds each. This might lead to heavy memory usage. We would recommend to comment out the lines that uses {\texttt{stim::write\_table\_data}} in the leakage/src/experiment.cpp and recompile before running the scripts for running leakage population evaluations. For LER evaluations, you need to uncomment these lines again and recompile before running the scripts. 

The script {\texttt{AE\_all\_data\_for\_plots\_wLER.sh}} runs LER evaluations and {\texttt{ AE\_all\_data\_for\_plots\_wo\_LER.sh}} runs other evaluations that doesn't require storing measurement data.

To run the data generation seperately for LER evaluations and leakage population evaluations, use the following,

\begin{highlightitembox}
\begin{verbatim}
$ cd build
$ cmake .. -DCMAKE_BUILD_TYPE=Release
$ make -j8   
$ cd ../leakage
$ ./AE_all_data_for_plots_wLER.sh <PROC>
\end{verbatim}
\end{highlightitembox}
To avoid unnecessary usage of memory, comment out the lines with {\texttt{stim::write\_table\_data}} from leakage/src/experiment.cpp, recompile and then run the other script.

\begin{highlightitembox}
\begin{verbatim}
$ make -j4
$ cd ../leakage
$ ./AE_all_data_for_plots_wo_LER.sh <PROC>
\end{verbatim}
\end{highlightitembox}

\subsection{Evaluation and expected results}
The results for LER, LRC and leakage population should be follow the 
same as the reported values in the paper, perhaps with slight deviations 
due to randomness.

\subsection{Experiment customization}
\oursm and \oursdm can be modified by modifying fleece.cpp file situated at quarch/src/fleece.cpp.
Similarly, experiments can be modified using leakage/src/experiments.cpp.

\subsection{Methodology}

Submission, reviewing and badging methodology:

\begin{itemize}
  \item \url{https://www.acm.org/publications/policies/artifact-review-and-badging-current}
  \item \url{https://cTuning.org/ae}
\end{itemize}



\section{Boolean Patterns for Leakage Detection}
\label{appendix:boolean_patterns}

To enable early identification of leaked qubits, we derive Boolean patterns based on detector outputs in various code configurations. These patterns are extracted from simulated leakage events and represent recurring structures in the detector graph that correlate with leakage. Below, we describe the methodology and present patterns for BPC codes, standard color codes, and the color code augmented with \oursdnspace.

\subsection{Pattern Extraction Methodology}
To derive accurate Boolean expressions for leakage patterns, we begin by identifying syndrome configurations associated with known leakage events using annotated simulations. Since these patterns vary in length depending on spatial and temporal detector context, we normalize them using an index tagging system that encodes shorter patterns with unique binary prefixes. For example, 6-bit patterns are padded to 7 bits with a leading \texttt{0}, 5-bit patterns with \texttt{10}, and so on, following a structured prefix scheme. This allows uniform processing while preserving distinctiveness between patterns of different lengths.

Once normalized, we construct a truth table over all possible bit combinations, marking those corresponding to known leakage signatures as true and the rest as false. Each true entry is then converted into a minterm, and symbolic Boolean minimization is applied using SymPy’s logic simplification engine. This produces compact, disjunctive normal form (DNF) expressions that generalize well across leakage events while maintaining logical equivalence. The final expressions are validated against the full truth table to ensure correctness and are integrated into our leakage speculation engine for efficient runtime detection.


\subsection{Balanced Product Cyclic Code Patterns}
The following Boolean expression encodes leakage patterns identified for the BPC code, as described in~\cite{kang2025quitsmodularqldpccode}.
\begin{small}
\begin{align*}
& (x_0 \wedge x_1 \wedge x_2 \wedge x_3 \wedge \neg x_5) \vee (x_0 \wedge x_1 \wedge x_2 \wedge x_4 \wedge \neg x_5) \\
& \vee (x_0 \wedge x_1 \wedge x_2 \wedge x_5 \wedge \neg x_6) \vee (x_0 \wedge x_1 \wedge x_3 \wedge x_4 \wedge \neg x_5) \\
& \vee (x_0 \wedge x_1 \wedge x_3 \wedge x_5 \wedge \neg x_6) \vee (x_0 \wedge x_1 \wedge x_4 \wedge x_5 \wedge \neg x_6) \\
& \vee (x_0 \wedge x_2 \wedge x_3 \wedge x_4 \wedge \neg x_5) \vee (x_1 \wedge x_2 \wedge x_3 \wedge x_4 \wedge \neg x_5) \\
& \vee (x_0 \wedge x_2 \wedge x_3 \wedge x_5 \wedge \neg x_4 \wedge \neg x_6) \vee (x_0 \wedge x_2 \wedge x_4 \wedge x_5 \wedge \neg x_3 \wedge \neg x_6) \\
& \vee (x_0 \wedge x_3 \wedge x_4 \wedge x_5 \wedge \neg x_2 \wedge \neg x_6) \vee (x_1 \wedge x_2 \wedge x_3 \wedge x_5 \wedge \neg x_4 \wedge \neg x_6) \\
& \vee (x_1 \wedge x_2 \wedge x_4 \wedge x_5 \wedge \neg x_3 \wedge \neg x_6) \vee (x_1 \wedge x_3 \wedge x_4 \wedge x_5 \wedge \neg x_2 \wedge \neg x_6)
\end{align*}
\end{small}

\subsection{Color Code Patterns}
Leakage detection patterns for the standard color code are more compact and are given below:
\begin{small}
\begin{align*}
& & (x_0 \wedge x_1 \wedge \neg x_2 \wedge \neg x_3) \vee (x_0 \wedge x_2 \wedge \neg x_1 \wedge \neg x_3) \vee (x_1 \wedge x_2 \wedge \neg x_0 \wedge \neg x_3)
\end{align*}
\end{small}

\subsection{Color Code with \oursd}
The following patterns were identified in the color code augmented with our speculation method (\oursdnspace), which captures more temporally distributed signatures of leakage:
\begin{small}
\begin{align*}
& (x_0 \wedge x_2 \wedge x_5 \wedge \neg x_1 \wedge \neg x_3 \wedge \neg x_6) \vee (x_0 \wedge x_2 \wedge x_5 \wedge \neg x_1 \wedge \neg x_4 \wedge \neg x_6) \\
& \vee (x_1 \wedge x_2 \wedge x_5 \wedge \neg x_0 \wedge \neg x_3 \wedge \neg x_6) \vee (x_1 \wedge x_2 \wedge x_5 \wedge \neg x_0 \wedge \neg x_4 \wedge \neg x_6) \\
& \vee (x_0 \wedge x_1 \wedge x_3 \wedge x_4 \wedge \neg x_2 \wedge \neg x_5 \wedge \neg x_6)\\ 
& \vee (x_0 \wedge x_1 \wedge x_3 \wedge x_5 \wedge \neg x_2 \wedge \neg x_4 \wedge \neg x_6) \\
& \vee (x_0 \wedge x_1 \wedge x_4 \wedge x_5 \wedge \neg x_2 \wedge \neg x_3 \wedge \neg x_6) \\
& \vee (x_0 \wedge x_2 \wedge x_3 \wedge x_4 \wedge \neg x_1 \wedge \neg x_5 \wedge \neg x_6) \\
& \vee (x_1 \wedge x_2 \wedge x_3 \wedge x_4 \wedge \neg x_0 \wedge \neg x_5 \wedge \neg x_6)
\end{align*}
\end{small}

\end{document}